\documentclass[journal]{IEEEtran}

\usepackage[colorlinks]{hyperref}
\usepackage{booktabs}
\usepackage{tabularx}
\usepackage{algorithm}
\usepackage{amssymb}
\usepackage{amsbsy}
\usepackage{amsmath}
\usepackage{bm}
\usepackage{verbatim}
\usepackage{mathrsfs}
\usepackage{amsfonts}
\usepackage{graphicx}
\usepackage[tight,footnotesize]{subfigure}
\usepackage[10pt]{moresize}
\usepackage{array}
\usepackage{color}
\usepackage{epsfig}
\usepackage{stfloats}
\usepackage{balance}
\usepackage{cancel}
\usepackage[noend]{algpseudocode}
\usepackage{algorithmicx}
\usepackage{setspace}
\usepackage{cases}
\usepackage{epstopdf}
\usepackage{multirow}
\usepackage{extarrows}
\usepackage{pifont}
\usepackage{enumerate}
\usepackage{enumitem}
\usepackage{titlesec}
\usepackage{cite}

\usepackage{placeins}

\usepackage{array}
\usepackage{booktabs}
\usepackage{multirow}
\usepackage{makecell}
\usepackage[normalem]{ulem} 

\titlespacing{\section}{0pt}{2 ex plus .0ex minus .0ex}{1ex plus .0ex}
\titlespacing{\subsection}{0pt}{1.5 ex plus .0ex minus .0ex}{0.8 ex plus 0.0ex}
\titlespacing{\subsubsection}{0pt}{0.5ex plus .0ex minus .0ex}{0.0ex plus .0ex}

\allowdisplaybreaks[4]

\begin{document}

\title{Joint Scheduling of Sensing Data Offloading and Edge Inference for Multi-UAV Networks}


\author{Yanan~Du, Sai~Xu, Yinbo~Yu

\vspace{-3mm}

\thanks{Y.~Du is with the Department of Electronic and Electrical Engineering, University of Sheffield, Sheffield, S1 4ET, UK (e-mail: $\rm yanan.du@ieee.org$). S. Xu is with University College London, London, UK (e-mail: $\rm sai.xu@ucl.ac.uk$). Y.~Yu is with College of Artificial Intelligence, Nanjing University of Aeronautics and Astronautics, Nanjing, Jiangsu, 210016, China (e-mail: $\rm yinboyu@nuaa.edu.cn$)}
}

\maketitle

\begin{abstract}
Unmanned aerial vehicles (UAVs) often collaborate by collecting and offloading sensing streams to an edge server, where a deep neural network (DNN) model performs cross-stream alignment, fusion, and inference. However, the coupling between wireless offloading and DNN execution makes end-to-end latency minimization challenging. To address this issue, this paper investigates efficient edge inference in multi-UAV networks. Specifically, a multi-UAV collaborative edge inference model is first established, in which UAV sensing streams are processed by a multi-branch DNN on a multi-core accelerator. Based on this model, an end-to-end latency minimization problem with a synchronization penalty is formulated. A genetic algorithm (GA)-based full joint scheduler, termed \texttt{GA-Joint}, is then developed to obtain high-quality scheduling solutions. To reduce the search complexity, two lightweight variants, termed \texttt{GA-DAG} and \texttt{GA-DACS}, are further proposed. Simulation results demonstrate that the proposed GA-based scheduling algorithms achieve lower end-to-end latency than \texttt{Decoupled-Greedy} and \texttt{Joint-Greedy}, which represent decoupled and joint greedy scheduling schemes, respectively, in most cases. Furthermore, \texttt{GA-DACS} achieves performance close to that of \texttt{GA-Joint} in many cases and even delivers slightly lower latency in certain scenarios.
\end{abstract}


\IEEEpeerreviewmaketitle

\section{Introduction}\label{sec:introduction}
\IEEEPARstart{U}{nmanned} aerial vehicles (UAVs) are increasingly networked to enable collaborative perception, supporting diverse intelligent applications such as inspection, surveillance, search and rescue, and infrastructure monitoring~\cite{Tang2025MUL-VR,Tian2025UCDNet}. Compared with single-UAV perception, multi-UAV collaboration provides richer observations, thereby alleviating occlusion, viewpoint limitations, and local information loss. Moreover, multi-UAV networks facilitate the acquisition and integration of complementary multimodal information, including RGB images, depth maps, LiDAR point clouds, and environmental cues, thereby improving perception robustness in complex operating environments~\cite{Feng2024U2UData,Wang2025UAVScenes,Hu2025ResourceAllocation}. However, achieving such information fusion often requires sensing streams from different UAVs to be offloaded over wireless uplinks to an edge server, where multi-branch deep neural network (DNN) models process heterogeneous inputs and perform feature-level or decision-level fusion for UAV perception tasks~\cite{Zhao2025MMFDet,Gu2025DEGFYOLO, You2026UAVMEC, Ozer2023Offloading, Zeng2023MEC}.

Under this architecture, the end-to-end latency of a single perception process is determined not only by sensing data offloading and downstream DNN execution, but also by their cross-stage interaction. In particular, asynchronous branch execution may provide latency benefits over a strictly synchronous strategy, as an early-arriving branch-specific sensing stream allows its corresponding DNN branch to start before the remaining streams complete transmission. This creates an opportunity to exploit cross-stage parallelism between communication and computation. 

However, exploiting such parallelism is nontrivial because the system exhibits strong cross-stage coupling. On the communication side, wireless resources are limited and the channel conditions of different UAVs vary over time~\cite{Dong2025Blockchain,Feng2021Hybrid}, requiring the scheduler to determine which sensing streams should be prioritized at each instant. On the computation side, edge resources are also constrained, and the resulting inference latency depends on operator precedence constraints, core assignment, and memory access overhead among directed acyclic graph (DAG) nodes~\cite{MAESTRO,Magma}. More importantly, the communication schedule directly determines the release times of branch-entry nodes, while the DAG structure and fusion dependencies, in turn, affect the urgency of uploading different UAV streams. Therefore, optimizing communication and computation in isolation may fail to fully reduce end-to-end latency. Despite its importance, the cross-stage coupling between sensing data offloading and DNN inference execution in multi-UAV systems remains underexplored. Motivated by these observations, this paper focuses on the joint scheduling of sensing data offloading and edge inference for multi-UAV networks.

\subsection{Related Work}
The problem of sensing data offloading and edge inference in multi-UAV networks is closely related to edge inference and DNN execution on hardware platforms. 
Although these two aspects are inherently coupled in practical systems, they have often been investigated separately. This section reviews these two lines of work and motivates their joint optimization.

\textit{1) Edge Inference:}
Edge inference has been extensively studied from the perspective of service-oriented computation, where inference requests are commonly abstracted as workloads to be dispatched, partitioned, or offloaded across distributed computing resources. Such a formulation is effective for capturing device--edge cooperation, but it often hides the fine-grained execution behavior of DNN operators on accelerators, as well as hardware-dependent bottlenecks such as limited on-chip memory, external memory access, and bandwidth contention. A representative paradigm in this area is collaborative inference, which aims to distribute DNN execution between resource-constrained end devices and more capable edge or cloud servers~\cite{Zeng2020Edge}. Existing methods typically determine a partition point in the network or computation graph, such that early-stage computations are performed locally and the remaining subnetwork is executed remotely. In this setting, the data sent over the uplink is no longer the original input but the intermediate representation produced by the local subnetwork~\cite{Yang2020Energy,Li2023Throughput}. Since the volume of such representations can still be substantial, various compression-oriented approaches have been introduced to reduce the amount of transmitted information and thus improve the latency and bandwidth efficiency of edge inference~\cite{Shao2021Learning, FrankenSplit2024Furutanpey}. Beyond representation compression, task-oriented communication further reformulates the objective of transmission. Instead of preserving all signal details, it seeks to retain only the semantic or task-discriminative information required by the downstream inference task~\cite{Li2025Tackling, Tarimo2026Adaptable,Xie2025Toward}. These studies demonstrate the potential of communication-efficient edge inference, but they usually do not explicitly model how the offloaded representations affect accelerator-level execution on the target hardware.

\textit{2) DNN Execution:}
Another line of work studies DNN inference from the perspective of hardware execution, with a particular focus on scheduling heterogeneous workloads on multi-core accelerators. Unlike service-level inference optimization, accelerator-level execution must account for dependencies among DNN operators, the mapping of operators to processing cores, memory allocation, and runtime resource contention. This leads to a large and highly coupled scheduling space, where both the temporal execution order and the spatial computation assignment need to be optimized. To make this problem tractable, prior studies have explored heuristic, search-based, and learning-based scheduling mechanisms. For data-center inference, AI-MT~\cite{AI-MT} designed a layer-granularity scheduling policy for concurrent DNN execution. MAGMA~\cite{Magma} and COMB~\cite{COMB} adopted evolutionary search to explore scheduling decisions more efficiently, with COMB further incorporating memory-related constraints into the optimization process. MoCA~\cite{MoCA} addressed multi-DNN execution by improving runtime memory management. In edge-oriented accelerator systems, Herald~\cite{Herald} and DREAM~\cite{DREAM} focused on meeting real-time inference requirements under constrained resources: the former exploited heterogeneous dataflow execution, whereas the latter relied on adaptive online scheduling to handle workload dynamics. DySta~\cite{DySta} considered sparse DNN workload scheduling, and TaiChi~\cite{TaiChi} combined graph neural networks with reinforcement learning to guide scheduling decisions on multi-core accelerators. While these studies improve hardware execution efficiency, they generally take DNN inputs as given and do not jointly consider how upstream wireless offloading decisions affect the release times of DNN operators on the accelerator.


Treating edge inference and accelerator-level execution as two isolated optimization problems may lead to suboptimal end-to-end performance. This motivates our cross-stage formulation, which jointly captures the communication behavior of sensing data offloading and the execution dynamics of DNN workloads on hardware platforms.

\subsection{Contributions}
In multi-UAV networks that offload sensing data to an edge server for inference, end-to-end latency is jointly determined by sensing data offloading and DNN execution on the edge hardware platform. These two stages can partially overlap: once a sensing stream arrives at the edge server, its corresponding DNN branch can immediately start execution. To exploit this inter-stage parallelism and hide part of the communication latency behind edge computation, this paper jointly optimizes sensing data offloading and edge inference to minimize end-to-end latency. The main contributions are summarized as follows.

\begin{itemize}[leftmargin=*]
\item A unified communication-computation model is established to characterize sensing data offloading and edge inference in multi-UAV networks, capturing the coupling between wireless communication and DNN execution on a multi-core accelerator. Based on this model, an end-to-end latency minimization problem is formulated by explicitly incorporating a synchronization penalty.

\item A genetic algorithm (GA)-based full joint scheduler, termed \texttt{GA-Joint}, is developed to jointly optimize uplink resource allocation and DNN execution on the multi-core accelerator. In \texttt{GA-Joint}, DAG criticality, synchronization urgency, and core-locality features are embedded into communication and computation decisions, enabling GA to search over scheduling policies rather than raw schedules. 

\item To reduce computational complexity, two lightweight variants are further designed. Specifically, \texttt{GA-DACS} optimizes a DAG-aware communication scheduling policy while using a greedy release-aware DAG scheduler, whereas \texttt{GA-DAG} fixes the communication scheduler and applies GA only to DAG priority assignment and core mapping.

\item Simulations are conducted to validate the effectiveness of the proposed algorithms through end-to-end execution timeline analysis and sensitivity studies with respect to the number of accelerator cores, the number of orthogonal frequency-division multiple access (OFDMA) subcarriers, the signal-to-interference-plus-noise ratio (SINR) threshold, communication load, and different branch-length configurations.
\end{itemize}

The remainder of this paper is organized as follows. Section~\ref{sec:system_model} introduces the system model and formulates the optimization problem. Section~\ref{sec:proposed_method} presents the proposed GA-based joint scheduling method. Section~\ref{sec:ga-dacs} describes two lightweight variants. Section~\ref{sec:simulation_results} presents the simulation settings and results. Section~\ref{sec:conclusions} concludes this paper.

\section{System Model}\label{sec:system_model}
\begin{figure}
\centering
\includegraphics[width=\linewidth]{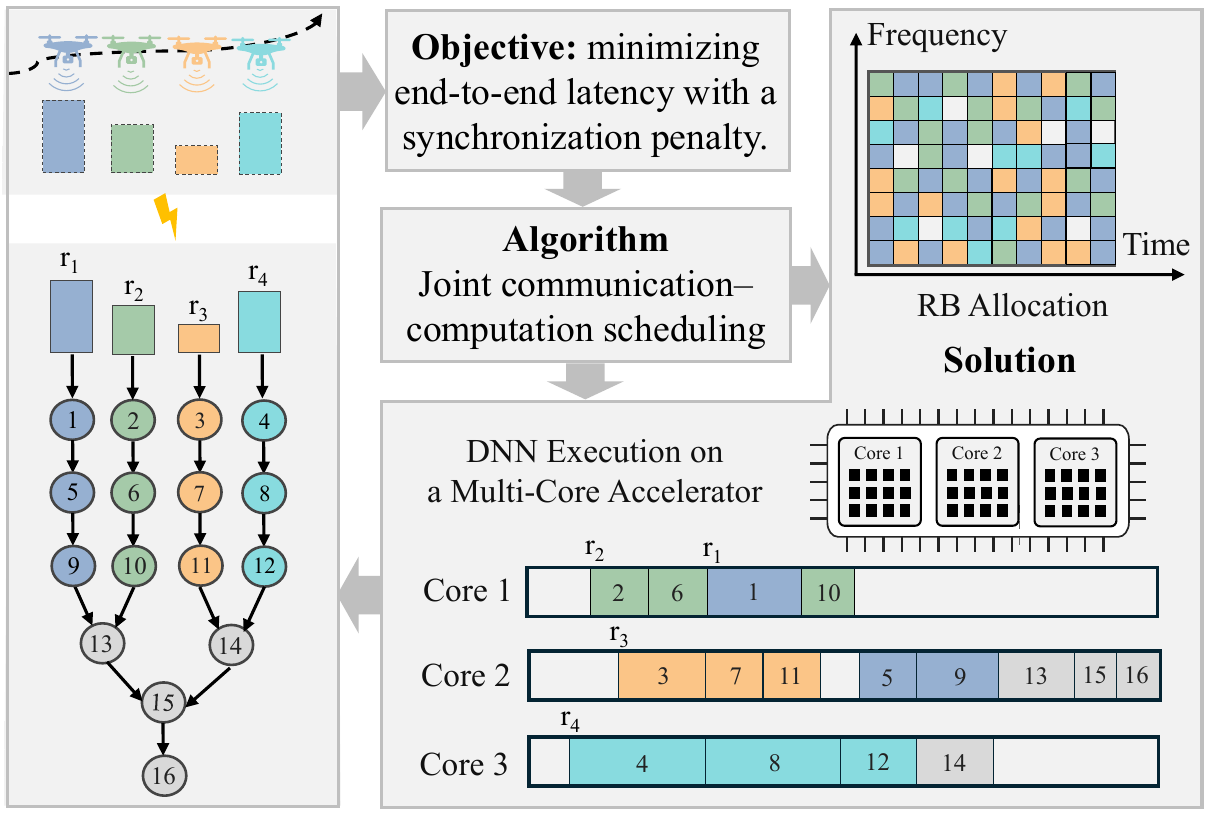}
\caption{An illustration of the multi-UAV collaborative edge inference system.}\label{Fig1}
\end{figure}
Fig. \ref{Fig1} illustrates a multi-UAV collaborative edge inference system, where multiple UAVs follow pre-planned trajectories, continuously collect sensing data, and upload their sensing data streams to an edge server via wireless uplinks. Each UAV is assumed to be responsible for a single sensing modality, establishing a one-to-one correspondence between UAVs and sensing modalities. The received data are then fed into a multi-branch DNN for collaborative perception, in which each sensing stream is processed by a dedicated branch. The DNN is executed on a multi-core accelerator.

\subsection{Communication Model}
During system operation, the sensing streams indexed by \(k\in \mathcal{K}=\{1,2,\ldots,K\}\) are transmitted to the edge server over uplink OFDMA. A one-to-one correspondence is assumed between UAVs, sensing modalities, and sensing streams. The transmission process is divided into discrete time slots indexed by \(t\in \mathcal{T}=\{1,2,\ldots,T\}\). In each slot, the available spectrum is divided into \(F\) orthogonal subcarriers, indexed by \(f\in \mathcal{F}=\{1,2,\ldots,F\}\). Each time-slot--subcarrier pair constitutes a resource block (RB). Each RB can be allocated to at most one sensing stream, while a sensing stream is allowed to occupy multiple orthogonal subcarriers within the same time slot. This setting captures OFDMA-based parallel uplink transmission, where multiple RBs in the same slot may jointly serve the same stream.

Let \(x_{k,t,f}\in\{0,1\}\) denote the RB allocation variable, where \(x_{k,t,f}=1\) indicates that stream \(k\) is scheduled on RB \((t,f)\), and \(x_{k,t,f}=0\) otherwise. The per-RB exclusiveness constraint is expressed as
\begin{equation}
\sum_{k=1}^{K} x_{k,t,f} \le 1,\quad 
\forall t\in\mathcal{T},\ \forall f\in\mathcal{F}.
\label{eq:rb_constraint}
\end{equation}
No additional constraint is imposed on \(\sum_{f=1}^{F}x_{k,t,f}\), since the same stream may be assigned multiple subcarriers in one time slot.

Let \(\gamma_k(t,f)\) denote the predicted uplink SINR\footnote{Since the UAV trajectories are pre-planned, the uplink SINR evolution in each collaborative inference round is assumed to be predictable over the scheduling horizon. Hence, the edge scheduler can exploit the predicted SINR values of different UAVs in future slots to perform forward-looking resource allocation during the current round.} of stream \(k\) on RB \((t,f)\), and let the threshold value \(\gamma_{\mathrm{th}}\) denote the minimum SINR required for successful transmission. If \(\gamma_k(t,f)<\gamma_{\mathrm{th}}\), 
then stream \(k\) is infeasible for transmission on the corresponding RB. Otherwise, its spectral efficiency is modeled using the clipped Shannon formula
\begin{equation}
\eta_k(t,f)=
\min\left\{
\log_2\!\left(1+\frac{\gamma_k(t,f)}{\Gamma_{\mathrm{gap}}}\right),
\eta_{\max}
\right\},
\label{eq:spectral_efficiency}
\end{equation}
where \(\Gamma_{\mathrm{gap}}\) is the Shannon-gap factor and \(\eta_{\max}\) is the maximum spectral efficiency. The achievable data rate of stream \(k\) on RB \((t,f)\) is then given by
\begin{equation}
R_k(t,f)=B_{\mathrm{RB}}\,\eta_k(t,f),
\label{eq:rate}
\end{equation}
where \(B_{\mathrm{RB}}\) denotes the bandwidth of one RB. If \(\gamma_k(t,f)<\gamma_{\mathrm{th}}\), then \(R_k(t,f)=0\).

In each collaborative inference round, stream \(k\) has an initial payload size \(D_k\). 
Since RBs are allocated sequentially in the proposed decoder, the payload evolution is modeled at the RB level. 
Let \(Q_k(t,f)\) denote the remaining payload of stream \(k\) immediately before processing RB \((t,f)\), with \(Q_k(1,1)=D_k\). 
After RB \((t,f)\) is processed, the remaining payload is updated as
\begin{equation}
\widetilde{Q}_k(t,f)
=
\max\left\{
0,\,
Q_k(t,f)-x_{k,t,f}R_k(t,f)\Delta t
\right\},
\label{eq:queue_rb_update}
\end{equation}
where \(\widetilde{Q}_k(t,f)\) denotes the post-allocation payload state. 
The queue state is propagated across RBs as
\begin{equation}
Q_k(t,f+1)=\widetilde{Q}_k(t,f),\quad 
\ f=1,\ldots,F-1,
\label{eq:queue_next_subcarrier}
\end{equation}
and
\begin{equation}
Q_k(t+1,1)=\widetilde{Q}_k(t,F),\quad 
t=1,\ldots,T-1.
\label{eq:queue_next_slot}
\end{equation}
Accordingly, the upload completion slot of stream \(k\) is defined as
\begin{equation}
\tau_k
=
\min\left\{
t\in\mathcal{T}
\,\middle|\,
\exists f\in\mathcal{F},\ \widetilde{Q}_k(t,f)=0
\right\}.
\label{eq:arrival}
\end{equation}
The completion slot \(\tau_k\) determines the release time of the corresponding branch-entry node for downstream DNN execution. 
Therefore, the communication schedule governs not only the upload completion time of each stream, but also the time at which the associated inference branch becomes available at the edge server.

\subsection{Multi-Branch DAG Inference and Execution Model}

The DNN executed at the edge server is modeled as a DAG \(G=(\mathcal{V},\mathcal{E})\), where each node \(v\in\mathcal{V}\) denotes a computational operator and each edge \((u,v)\in\mathcal{E}\) represents a precedence dependency. The DAG contains three classes of nodes: branch-specific nodes that form the per-stream computation chains, group-alignment nodes that aggregate selected branches according to a predefined grouping, and shared fusion-head nodes that process the aligned features through a common fusion head, classifier, and output chain. Hence, the DAG captures both intra-branch dependencies and inter-branch aggregation dependencies. Consistent with the communication model, each input sensing stream corresponds to one DAG branch. Let \(r_k\in\mathcal{V}\) denote the branch-entry node corresponding to stream \(k\in\mathcal{K}\). Since branch \(k\) becomes executable only after stream \(k\) has been fully uploaded, the release time of \(r_k\) is determined by the upload completion time \(\tau_k \Delta t\). Accordingly, the release time of node \(v\) is defined as
\begin{equation}
\rho_v=
\begin{cases}
\tau_k \Delta t, & \text{if } v=r_k,\ \forall k\in\mathcal{K},\\
-\infty, & \text{otherwise.}
\end{cases}
\label{eq:node_release}
\end{equation}

At the edge server, DNN inference is executed on a multi-core accelerator consisting of \(C\) homogeneous cores, indexed by \(\mathcal{C}=\{1,\ldots,C\}\). For node \(v\), 
\(a_v\in\mathcal{C}\) denotes its assigned core, while \(p_v\), \(s_v\), and \(f_v\) denote its computation time, start time, and finish time, respectively. In addition, 
\(t_v^{\mathrm{r,on}}\), \(t_v^{\mathrm{w,on}}\), \(t_v^{\mathrm{r,off}}\), and \(t_v^{\mathrm{w,off}}\) denote its on-chip read time, on-chip write time, off-chip read 
time, and off-chip write time, respectively. For each dependency edge \((u,v)\in\mathcal{E}\), \(\Phi_{u\rightarrow v}\) denotes the time at which the output of node \(u\) becomes available to node \(v\). If the output tensor of \(u\) can be retained in the local on-chip buffer of the core assigned to \(v\), on-chip reuse is exploited, and
\begin{equation}
\Phi_{u\rightarrow v}=f_u+t_u^{\mathrm{w,on}}+t_v^{\mathrm{r,on}}.
\label{eq:onchip_transfer}
\end{equation}
Otherwise, off-chip transfer is required, and
\begin{equation}
\Phi_{u\rightarrow v}=f_u+t_u^{\mathrm{w,off}}+t_v^{\mathrm{r,off}}.
\label{eq:offchip_transfer}
\end{equation}

Let \(\chi_c\) denote the availability time of core \(c\). Then, the start time of node \(v\) assigned to core \(a_v\) must satisfy
\begin{equation}
s_v\ge
\max\left\{
\rho_v,\,
\chi_{a_v},\,
\max_{u\in\mathrm{Pred}(v)} \Phi_{u\rightarrow v}
\right\},
\quad \forall v\in\mathcal{V},
\label{eq:start_constraint}
\end{equation}
where \(\mathrm{Pred}(v)\) denotes the set of predecessor nodes of \(v\). The finish time of node \(v\) is then
\begin{equation}
f_v=s_v+p_v.
\label{eq:finish_time}
\end{equation}
In addition, each core can execute at most one node at a time. Therefore, any two distinct nodes assigned to the same core must satisfy
\begin{equation}
[s_{v'},f_{v'})\cap[s_v,f_v)=\varnothing,
\quad \forall v'\neq v,\ \text{if } a_{v'}=a_v.
\label{eq:nonoverlap}
\end{equation}

\subsection{Synchronization-Aware Fusion Model}
The considered multi-branch DAG contains group-wise alignment and fusion modules, where input streams associated with the same target or region are processed jointly before entering the shared inference head. As a result, the efficiency of downstream fusion depends not only on whether each individual stream is uploaded successfully, but also on how well the arrival times of correlated streams are aligned. Additionally, such synchronization affects the buffering overhead and the semantic consistency of multi-stream features at the edge server, both of which are critical to reliable collaborative inference.

Let \(\mathcal{G}=\{1,\ldots,M\}\) denote the set of input groups, and \(g(k)\in\mathcal{G}\) the group index of stream \(k\). For each group \(m\in\mathcal{G}\), define the synchronization mismatch as the spread of upload completion times within that group, i.e.,
\begin{equation}
\Delta_m=\max_{k:g(k)=m}\tau_k \Delta t -\min_{k:g(k)=m}\tau_k \Delta t.
\label{eq:group_sync}
\end{equation}
A larger \(\Delta_m\) indicates a more severe temporal mismatch among the sensing streams in group \(m\), which may delay the corresponding alignment and fusion operations at the edge server.
Based on this, the total synchronization penalty is defined as
\begin{equation}
P_{\mathrm{sync}}=\sum_{m\in\mathcal{G}}\Delta_m.
\label{eq:sync_penalty}
\end{equation}
This term is introduced to explicitly account for inter-stream temporal coordination in collaborative inference. Without such a term, the scheduler may over-prioritize individually favorable transmissions while neglecting the arrival alignment among correlated sensing streams, which can still delay downstream fusion even if some streams are delivered early. In addition to latency, such temporal coordination also helps reduce unnecessary buffering at the edge server and preserve the semantic consistency of multi-stream features. 

\subsection{Problem Formulation}
The goal of the considered system is to coordinate uplink transmission and edge-side DNN execution so that all sensing streams are delivered in a timely manner and the 
final collaborative inference result is produced as early as possible. Define the end-to-end latency as the completion time of the overall DNN inference process, which is 
equivalently given by the finish time of the last DAG node, i.e.,
\begin{equation}
T_{\mathrm{e2e}} = \max_{v\in\mathcal{V}} f_v.
\label{eq:e2e}
\end{equation}
However, minimizing only \(T_{\mathrm{e2e}}\) is not sufficient for multi-UAV collaborative perception. To mitigate the additional waiting time, buffering overhead, and feature inconsistency caused by synchronization mismatch, a synchronization penalty is explicitly incorporated into the overall latency through the following composite objective:
\begin{equation}
J = T_{\mathrm{e2e}} + \lambda P_{\mathrm{sync}},
\label{eq:objective}
\end{equation}
where \(\lambda \ge 0\) is a tunable weighting factor that balances end-to-end latency reduction and temporal coordination among grouped sensing streams. Note that although fusion dependencies already reflect part of the waiting effect in $T_{\mathrm{e2e}} $, $P_{\mathrm{sync}}$ is introduced to explicitly regularize inter-stream temporal alignment and buffering/semantic consistency, which are not fully captured by the final completion time alone.

Taking these factors into account, the joint communication--computation scheduling problem can be formulated as
\begin{equation}
\begin{aligned}
\min_{\substack{\{x_{k,t,f}\}\\ \{a_v,s_v,f_v\}}} \quad
& T_{\mathrm{e2e}} + \lambda P_{\mathrm{sync}} \\
\mathrm{s.t.}\quad
& x_{k,t,f}\in\{0,1\},~ 
\forall k\in\mathcal{K},~ t\in\mathcal{T},~ f\in\mathcal{F}, \\
& x_{k,t,f}=0,\quad 
\text{if } \gamma_k(t,f)<\gamma_{\mathrm{th}},\ 
\forall k,t,f, \\
& Q_k(1,1)=D_k,\quad \forall k\in\mathcal{K}, \\
& a_v\in\mathcal{C},\quad \forall v\in\mathcal{V}, \\
& s_v\ge 0,\quad f_v\ge 0,\quad \forall v\in\mathcal{V}, \\
& \eqref{eq:rb_constraint},\ 
\eqref{eq:queue_rb_update} - \eqref{eq:node_release}, \ \eqref{eq:start_constraint},\ 
\eqref{eq:finish_time},\ 
\eqref{eq:nonoverlap}.
\end{aligned}
\label{eq:main_problem}
\end{equation}
In problem \eqref{eq:main_problem}, the binary variables \(\{x_{k,t,f}\}\) specify the OFDMA uplink scheduling decisions over each RB, while \(\{a_v,s_v,f_v\}\) 
characterize the core assignment and execution timing of DAG nodes at the multi-core accelerator.

\section{GA-Based Full Joint Scheduling (\texttt{GA-Joint})}
\label{sec:proposed_method}

The optimization problem in \eqref{eq:main_problem} is challenging to solve directly due to the strong coupling between communication and computation. Specifically, the uplink 
scheduling decisions \(\{x_{k,t,f}\}\) determine the upload completion times \(\{\tau_k\}\), which define the release times of the branch-entry DAG nodes and 
consequently influence the subsequent execution order and core-mapping decisions on the multi-core accelerator. In addition, since multiple streams within the same group are 
jointly aligned and fused in downstream processing, their relative completion times also affect the resulting synchronization penalty.

Directly optimizing all slot-level and node-level decisions is computationally prohibitive because the resulting search space is highly combinatorial. To address this difficulty, 
this paper adopts a parameterized policy-search approach for GA-based full joint scheduling, namely \texttt{GA-Joint}. Instead of searching over raw schedules, a candidate scheduler is 
represented by a compact real-valued vector, from which a complete communication--computation schedule is deterministically decoded. The decoded schedule is then evaluated 
using the objective in \eqref{eq:objective}, and the policy vector is optimized by a GA. This design preserves the main coupling structure of the original problem while keeping 
the search space manageable.

\subsection{Parameterized Joint Scheduling Policy}

\texttt{GA-Joint} represents a candidate joint scheduler by a 20-dimensional real-valued policy vector
\begin{equation}
\boldsymbol{\theta}=
\big[\bm{\alpha},\bm{\beta},\bm{\mu}\big]
\in \mathbb{R}^{20},
\label{eq:policy_vector}
\end{equation}
where
\begin{equation}
\bm{\alpha}\in\mathbb{R}^{8},\qquad
\bm{\beta}\in\mathbb{R}^{8},\qquad
\bm{\mu}\in\mathbb{R}^{4}.
\label{eq:policy_partition}
\end{equation}
Here, \(\bm{\alpha}\) controls the communication-stage stream selection rule, \(\bm{\beta}\) controls the ready-node priority rule during DAG scheduling, and 
\(\bm{\mu}\) controls the core-mapping rule. In the communication stage, the scheduler decides which stream should occupy each RB \((t,f)\). Once the upload completion times are 
determined, the release times of the corresponding branch-entry nodes are fixed. The computation stage then repeatedly selects one ready node and maps it onto one of the 
available cores. Therefore, rather than directly encoding \(\{x_{k,t,f}\}\), \(\{a_v\}\), and \(\{s_v,f_v\}\), the policy vector encodes only the decision rules used to generate 
them. For clarity, the interpretation of all entries in the policy vector is summarized in Table~\ref{tab:weight_interpretation}.

\begin{table*}[t]
\centering
\scriptsize
\caption{Interpretation of the 20-Dimensional Policy Vector.}
\label{tab:weight_interpretation}
\renewcommand{\arraystretch}{1.15}
\setlength{\tabcolsep}{5pt}
\begin{tabular}{c|c|p{4.6cm}|p{9.8cm}}
\hline
\textbf{Entry} & \textbf{Stage} & \textbf{Associated Feature} & \textbf{Interpretation} \\
\hline
$\alpha_1$  & Comm. & Remaining payload $Q_k(t,f)$ & Controls whether sensing streams with larger unfinished data volumes should be prioritized. \\
$\alpha_2$  & Comm. & SINR $\gamma_k(t,f)$ & Measures the preference for favorable instantaneous channel quality. \\
$\alpha_3$  & Comm. & Rate $R_k(t,f)$ & Reflects the preference for immediate transmission gain on the current RB. \\
$\alpha_4$  & Comm. & Branch length & Prioritizes sensing streams associated with longer downstream computation branches. \\
$\alpha_5$  & Comm. & Unfinished-stream count in the same group & Encodes group-level urgency from the synchronization perspective. \\
$\alpha_6$  & Comm. & Finished-stream count in the same group & Captures whether the current stream should be accelerated to improve intra-group alignment. \\
$\alpha_7$  & Comm. & Slot index $t$ & Allows the communication policy to adapt across different scheduling phases. \\
$\alpha_8$  & Comm. & Inverse rate & Captures the penalty associated with low-rate transmission opportunities. \\
\hline
$\beta_1$  & DAG & Bottom level & Controls how strongly critical-path nodes are prioritized. \\
$\beta_2$  & DAG & Computation time $p_v$ & Reflects the preference for nodes with longer execution times. \\
$\beta_3$  & DAG & Number of successors & Emphasizes nodes that can release more downstream tasks after completion. \\
$\beta_4$  & DAG & Negative predecessor count & Penalizes nodes with more complex dependency structures. \\
$\beta_5$  & DAG & Alignment/fusion-chain indicator & Highlights nodes that belong to the downstream alignment or fusion path. \\
$\beta_6$  & DAG & Cross-branch indicator & Highlights nodes responsible for inter-branch aggregation. \\
$\beta_7$  & DAG & Negative release time & Favors nodes that become executable earlier. \\
$\beta_8$  & DAG & Negative average core availability & Makes node priority adaptive to the current global compute-load state. \\
\hline
$\mu_1$ & Map & Cross-core dependency count & Penalizes candidate mappings that induce more cross-core predecessor transfers. \\
$\mu_2$ & Map & Off-chip access delay & Penalizes mappings with higher off-chip communication overhead. \\
$\mu_3$ & Map & Same-core predecessor count & Rewards locality-preserving placements that improve on-chip reuse. \\
$\mu_4$ & Map & Core availability time & Reflects sensitivity to core waiting time and load balancing. \\
\hline
\end{tabular}
\end{table*}

\subsection{Deterministic Joint Decoding}

Given a policy vector \(\boldsymbol{\theta}\), the decoder generates a complete solution in two coupled stages.

\subsubsection{Communication-Stage Decoding}

At each RB \((t,f)\), only streams with unfinished uploads and feasible uplink quality on that RB can be scheduled. The eligible set is defined as
\begin{equation}
\mathcal{A}(t,f)=
\left\{k\in\mathcal{K}\,\middle|\,
Q_k(t,f)>0,\ \gamma_k(t,f)\ge\gamma_{\mathrm{th}}
\right\}.
\label{eq:eligible_set}
\end{equation}
If \(\mathcal{A}(t,f)=\varnothing\), the corresponding RB remains idle. Otherwise, for each eligible stream \(k\in\mathcal{A}(t,f)\), an 8-dimensional feature vector is constructed as
\begin{equation}
\mathbf{z}_k(t,f)=\big[z_{k,1}(t,f),\ldots,z_{k,8}(t,f)\big],
\label{eq:comm_feature}
\end{equation}
whose entries respectively correspond to remaining payload, instantaneous SINR, achievable rate, branch length, unfinished-stream count in the same group, finished-stream count in 
the same group, slot index, and inverse rate. The communication priority score is then defined as
\begin{equation}
\Psi_k^{\mathrm{comm}}(t,f)=
\sum_{i=1}^{8}\alpha_i z_{k,i}(t,f).
\label{eq:comm_score}
\end{equation}
The scheduler selects the stream with the largest score on the current RB,
\begin{equation}
k_{t,f}^{\star}=
\arg\max_{k\in\mathcal{A}(t,f)}
\Psi_k^{\mathrm{comm}}(t,f).
\label{eq:comm_decision}
\end{equation}
After the selected stream is assigned to RB \((t,f)\), the payload evolution is updated according to \eqref{eq:queue_rb_update}--\eqref{eq:queue_next_slot}. Repeating this process across all RBs within the scheduling 
horizon yields the complete OFDMA upload schedule and the resulting completion times \(\{\tau_k\}\).

It is important to emphasize that this communication-stage decoder is feature-based. Unlike an exhaustive per-slot look-ahead scheduler, it does not invoke the DAG scheduler 
at every RB allocation step. Instead, the cross-stage coupling is captured implicitly through the policy optimization process, because the policy weights are learned according 
to the final end-to-end objective after full communication--computation decoding.

\subsubsection{Release-Time Propagation and DAG Scheduling}

Once the communication stage is completed, the upload completion times are translated into branch-entry release times. In this paper, the release time of node \(v\) is defined in \eqref{eq:node_release}.
Here, \(\rho_v=-\infty\) indicates that node \(v\) is not directly constrained by an external upload-completion time. Its actual start time is still determined by predecessor 
dependencies and core availability.

The DAG is then scheduled iteratively on the multi-core accelerator. Let \(\mathcal{R}\) denote the current ready-node set, i.e., the set of unscheduled nodes whose predecessors 
have all finished. The release-time constraint is enforced later in the start-time calculation through \(\rho_v\). For each ready node \(v\in\mathcal{R}\), define an 
8-dimensional feature vector
\begin{equation}
\mathbf{q}_v=\big[q_{v,1},\ldots,q_{v,8}\big],
\label{eq:dag_feature}
\end{equation}
where the adopted features are the bottom level, node computation time, number of successors, negative number of predecessors, indicator of whether the node belongs to the 
alignment/fusion chain, indicator of whether it is a cross-branch node, negative release time, and negative average core availability. 

The ready-node priority score is given by
\begin{equation}
\Psi_v^{\mathrm{dag}}=
\sum_{i=1}^{8}\beta_i q_{v,i}.
\label{eq:dag_score}
\end{equation}
The next scheduled node is selected as
\begin{equation}
v^{\star}=
\arg\max_{v\in\mathcal{R}}
\Psi_v^{\mathrm{dag}}.
\label{eq:dag_decision}
\end{equation}

After selecting \(v^{\star}\), the scheduler evaluates all candidate cores. If node \(v^{\star}\) is tentatively mapped to core \(c\in\mathcal{C}\), its candidate start time is
\begin{equation}
\hat{s}_{v^{\star},c}=
\max\left\{
\chi_c,\,
\rho_{v^{\star}},\,
\max_{u\in\mathrm{Pred}(v^{\star})}
\hat{\Phi}_{u\rightarrow v^{\star}}^{(c)}
\right\},
\label{eq:candidate_start}
\end{equation}
where \(\chi_c\) is the current availability time of core \(c\), \(\rho_{v^{\star}}\) is the release time of node \(v^{\star}\), and \(\hat{\Phi}_{u\rightarrow v^{\star}}^{(c)}\) 
is the predecessor-data ready time under the candidate core \(c\). For nodes without predecessors, the predecessor-related maximum is set to zero. The corresponding candidate finish time is
\begin{equation}
\hat{f}_{v^{\star},c}=
\hat{s}_{v^{\star},c}+p_{v^{\star}}.
\label{eq:candidate_finish}
\end{equation}
To balance latency and locality, define the mapping score
\begin{equation}
\Psi_{v^{\star},c}^{\mathrm{map}}=
\hat{f}_{v^{\star},c}
+\mu_1\kappa_{v^{\star},c}^{\mathrm{cross}}
+\mu_2\Delta_{v^{\star},c}^{\mathrm{off}}
-\mu_3\kappa_{v^{\star},c}^{\mathrm{same}}
+\mu_4 \chi_c,
\label{eq:mapping_score}
\end{equation}
where \(\kappa_{v^{\star},c}^{\mathrm{cross}}\) is the number of predecessor dependencies that require cross-core transfer, \(\Delta_{v^{\star},c}^{\mathrm{off}}\) is the off-chip 
access delay, and \(\kappa_{v^{\star},c}^{\mathrm{same}}\) is the number of same-core predecessor colocations. The selected core is
\begin{equation}
a_{v^{\star}}=
\arg\min_{c\in\mathcal{C}}
\Psi_{v^{\star},c}^{\mathrm{map}}.
\label{eq:mapping_decision}
\end{equation}
After the node is assigned, its start and finish times are updated, the core availability state is refreshed, and the ready set is updated. This process continues until all DAG 
nodes are completed.

\subsubsection{Fitness Evaluation}

The above decoding procedure converts a policy vector \(\boldsymbol{\theta}\) into a communication--computation schedule. The fitness value is defined as
\begin{equation}
F(\boldsymbol{\theta})=
T_{\mathrm{e2e}}(\boldsymbol{\theta})
+\lambda P_{\mathrm{sync}}(\boldsymbol{\theta})
+P_{\mathrm{inv}}(\boldsymbol{\theta}),
\label{eq:fitness}
\end{equation}
where \(P_{\mathrm{inv}}(\boldsymbol{\theta})\) is a large penalty used only when a candidate policy fails to complete all uploads within the scheduling horizon.

\subsection{GA-Based Policy Optimization}

Because the fitness function in \eqref{eq:fitness} is induced by discrete RB allocation, ready-node ordering, and core mapping, it is highly nonlinear and nondifferentiable with 
respect to the policy parameters. Therefore, a real-coded GA is adopted to optimize the policy vector \(\boldsymbol{\theta}\).

The role of the GA is to search over the continuous policy space defined by \eqref{eq:policy_vector}, rather than to directly enumerate the original combinatorial 
schedule space. For each chromosome, the deterministic decoder first constructs an OFDMA upload schedule and obtains the stream completion times \(\{\tau_k\}\). It then propagates 
the corresponding release times to the branch-entry nodes and generates a release-aware DAG execution schedule on the multi-core accelerator. The resulting end-to-end latency and 
synchronization penalty are subsequently combined according to \eqref{eq:fitness} to evaluate the chromosome. Based on the fitness values, the population is iteratively updated 
through selection, crossover, mutation, and elite preservation. In this way, the population progressively evolves toward policy vectors that yield lower latency and a lower 
synchronization penalty.

An important feature of this method is that, once all uploads are completed within the scheduling horizon, the decoder deterministically produces a schedule satisfying the 
communication and DAG execution constraints. Therefore, the GA mainly focuses on improving policy quality, while incomplete-upload cases are handled by the penalty term 
\(P_{\mathrm{inv}}\). The overall execution flow can be summarized as \emph{policy generation $\rightarrow$ schedule decoding $\rightarrow$ fitness evaluation 
$\rightarrow$ population update}. Algorithm~\ref{alg:ga_joint} details \texttt{GA-Joint}, including communication-stage decoding, release-time propagation, DAG scheduling, and fitness 
evaluation.

\algrenewcommand\algorithmicrequire{\textbf{Input:}}
\algrenewcommand\algorithmicensure{\textbf{Output:}}

\begin{algorithm}[!t]
\caption{GA-Based Full Joint Scheduling (\texttt{GA-Joint})}
\label{alg:ga_joint}
\begin{algorithmic}[1]
\Require $G=(\mathcal{V},\mathcal{E})$, $\mathcal{K}$, $\mathcal{F}$, $\mathcal{C}$, 
$\{D_k\}_{k\in\mathcal{K}}$, $\{\gamma_k(t,f)\}$, $T_{\max}$, $\lambda$, $P$, $G_{\max}$
\Ensure $\boldsymbol{\theta}^{\star},\ J^{\star}$
\Statex \textit{Convention:} If $\mathrm{Pred}(v)=\varnothing$, the predecessor-related maximum is set to $0$. \(Q_k\) denotes the remaining payload of stream \(k\) immediately before the current RB is processed.
\State Initialize population
$\mathcal{P}^{(0)}=\{\boldsymbol{\theta}^{(p,0)}\}_{p=1}^{P}$, where
$\boldsymbol{\theta}^{(p,0)}=[\bm{\alpha}^{(p,0)},\bm{\beta}^{(p,0)},\bm{\mu}^{(p,0)}]$
\label{ln:joint:init-pop}

\For{$g=0$ to $G_{\max}-1$}
    \For{$p=1$ to $P$}

        \State $Q_k \gets D_k,\ \tau_k^{(p,g)} \gets +\infty,\ \forall k\in\mathcal{K}$
        \State $x_{k,t,f}^{(p,g)}\gets 0,\ \forall k\in\mathcal{K},\ t\le T_{\max},\ f\in\mathcal{F}$
        \label{ln:joint:eval-init}

        \For{$t=1$ to $T_{\max}$}
            \label{ln:joint:comm-begin}
            \For{$f\in\mathcal{F}$}
                \State $\mathcal{A}(t,f)\gets
                \{k\in\mathcal{K}\mid Q_k>0,\ \gamma_k(t,f)\ge \gamma_{\mathrm{th}}\}$

                \If{$\mathcal{A}(t,f)\neq\varnothing$}
                    \State $k^\star \gets
                    \arg\max\limits_{k\in\mathcal{A}(t,f)}
                    \Psi_k^{\mathrm{comm}}(t,f;\boldsymbol{\theta}^{(p,g)})$
                    \label{ln:joint:comm-priority}

                    \State $x_{k^\star,t,f}^{(p,g)}\gets 1$

                    \State $Q_{k^\star}\gets
                    \max\{0,\ Q_{k^\star}-R_{k^\star}(t,f)\Delta t\}$

                    \If{$Q_{k^\star}=0 \ \wedge\ \tau_{k^\star}^{(p,g)}=+\infty$}
                        \State $\tau_{k^\star}^{(p,g)}\gets t$
                    \EndIf
                \EndIf
            \EndFor

            \If{$Q_k=0,\ \forall k\in\mathcal{K}$}
                \State \textbf{break}
            \EndIf
        \EndFor
        \label{ln:joint:comm-end}

        \If{$\exists k\in\mathcal{K}: \tau_k^{(p,g)}=+\infty$}
            \label{ln:joint:infeas-begin}
            \State $F(\boldsymbol{\theta}^{(p,g)})\gets P_{\mathrm{inv}}$
            \State \textbf{continue}
        \EndIf
        \label{ln:joint:infeas-end}

        \State $\rho_v^{(p,g)}\gets
        \begin{cases}
        \tau_k^{(p,g)} \Delta t, & \text{if } v=r_k \text{ for some } k\in\mathcal{K},\\
        -\infty, & \text{otherwise}
        \end{cases}$
        \label{ln:joint:release}

        \State $\chi_c\gets 0,\ \forall c\in\mathcal{C}$; \quad
        $\mathcal{U}\gets \mathcal{V}$
        \label{ln:joint:dag-init}

        \While{$\mathcal{U}\neq\varnothing$}
            \label{ln:joint:dag-begin}

            \State $\mathcal{R}\gets
            \{v\in\mathcal{U}\mid \mathrm{Pred}(v)\cap \mathcal{U}=\varnothing\}$

            \State $v^\star \gets
            \arg\max\limits_{v\in\mathcal{R}}
            \Psi_v^{\mathrm{dag}}(\boldsymbol{\theta}^{(p,g)})$
            \label{ln:joint:dag-node}

            \State $c^\star \gets
            \arg\min\limits_{c\in\mathcal{C}}
            \Psi_{v^\star,c}^{\mathrm{map}}(\boldsymbol{\theta}^{(p,g)})$
            \label{ln:joint:dag-core}

            \State $a_{v^\star}\gets c^\star$

            \State $s_{v^\star}\gets
            \max\Bigl\{
            \rho_{v^\star}^{(p,g)},
            \chi_{c^\star},
            \max\limits_{u\in \mathrm{Pred}(v^\star)}
            \hat{\Phi}_{u\to v^\star}^{(c^\star)}
            \Bigr\}$
            \label{ln:joint:dag-start}

            \State $f_{v^\star}\gets s_{v^\star}+p_{v^\star}$
            \State $\chi_{c^\star}\gets f_{v^\star}$
            \State $\mathcal{U}\gets \mathcal{U}\setminus \{v^\star\}$

        \EndWhile
        \label{ln:joint:dag-end}

        \State $T_{\mathrm{e2e}}^{(p,g)}\gets
        \max\limits_{v\in\mathcal{V}} f_v$
        \label{ln:joint:obj-begin}

        \State $P_{\mathrm{sync}}^{(p,g)}\gets
        \Delta t\sum\limits_{m=1}^{M}
        \left(
        \max\limits_{k:g(k)=m}\tau_k^{(p,g)}
        -
        \min\limits_{k:g(k)=m}\tau_k^{(p,g)}
        \right)$

        \State $F(\boldsymbol{\theta}^{(p,g)})\gets
        T_{\mathrm{e2e}}^{(p,g)}+\lambda P_{\mathrm{sync}}^{(p,g)}$
        \label{ln:joint:obj-end}

    \EndFor

    \State $\mathcal{P}_{\mathrm{elite}}^{(g)}\gets
    \mathrm{Elite}\bigl(\mathcal{P}^{(g)},F\bigr)$
    \label{ln:joint:ga-begin}

    \State $\mathcal{P}_{\mathrm{off}}^{(g)}
    \gets
    \mathrm{Mutation}\!\left(
    \mathrm{Crossover}\!\left(
    \mathrm{Selection}\!\left(\mathcal{P}^{(g)},F\right)
    \right)
    \right)$

    \State $\mathcal{P}^{(g+1)}
    \gets
    \mathcal{P}_{\mathrm{off}}^{(g)}
    \cup
    \mathcal{P}_{\mathrm{elite}}^{(g)}$

    \If{$\mathrm{Stop}\!\left(\mathcal{P}^{(g+1)},\mathcal{P}^{(g)}\right)=1$}
        \State \textbf{break}
    \EndIf
    \label{ln:joint:ga-end}

\EndFor

\State $\boldsymbol{\theta}^{\star} \gets
\arg\min\limits_{\boldsymbol{\theta}\in \cup_g \mathcal{P}^{(g)}} F(\boldsymbol{\theta})$

\State $J^{\star} \gets F(\boldsymbol{\theta}^{\star})$

\State \Return $\boldsymbol{\theta}^{\star},\ J^{\star}$

\end{algorithmic}
\end{algorithm}

\FloatBarrier

\subsection{Complexity Analysis}

Let \(K=|\mathcal{K}|\), \(T=|\mathcal{T}|\), \(F=|\mathcal{F}|\), \(N=|\mathcal{V}|\), \(E=|\mathcal{E}|\), and \(C=|\mathcal{C}|\) denote the numbers of 
streams, time slots, subcarriers, DAG nodes, DAG edges, and compute cores, respectively. The complexity of one chromosome evaluation is determined by the deterministic decoder. 
In the communication stage, each RB \((t,f)\) requires scoring at most \(K\) candidate streams. Hence, the OFDMA decoding complexity is upper bounded by \(\mathcal{O}(TFK)\). 
In the computation stage, ready-node construction and priority evaluation require at most \(\mathcal{O}(N^2)\) operations in a straightforward implementation. The core-selection 
step evaluates \(C\) candidate cores for each scheduled node, and the predecessor-related timing updates over all nodes involve all DAG edges once per candidate core, resulting in 
\(\mathcal{O}(CE)\) complexity. Therefore, the per-chromosome decoding complexity can be upper bounded by
\begin{equation}
\mathcal{O}\big(TFK+N^2+CE\big).
\label{eq:single_decode_complexity}
\end{equation}
Let \(P\) and \(G_{\max}\) denote the population size and the maximum number of generations, respectively. Since each chromosome requires one full decoding pass, the overall complexity 
of the proposed method is given by
\begin{equation}
\mathcal{O}\Big(PG_{\max}\big(TFK+N^2+CE\big)\Big).
\label{eq:total_complexity}
\end{equation}
It is worth noting that the search is conducted in a fixed 20-dimensional continuous parameter space rather than in the original combinatorial schedule space. Consequently, 
although the optimization is iterative, the proposed method remains computationally tractable for the problem scale considered in this paper.

\section{Lightweight GA-Based Joint Scheduling}
\label{sec:ga-dacs}

To further reduce the search complexity of the full joint scheduling problem, this paper proposes two lightweight scheduling algorithms, termed \texttt{GA-DACS} and \texttt{GA-DAG}. Unlike \texttt{GA-Joint}, 
which optimizes a 20-dimensional policy vector covering communication scheduling, ready-node priority, and core mapping, the two lightweight variants use GA to optimize only part of the joint 
scheduling policy. Specifically, \texttt{GA-DACS} optimizes an 8-dimensional communication-stage policy and adopts a greedy release-aware DAG scheduler for downstream execution. In 
contrast, \texttt{GA-DAG} uses a fixed greedy communication scheduler and optimizes a 12-dimensional DAG-stage policy, including ready-node priority and core-mapping rules.

Although the two lightweight methods reduce the chromosome dimension and search space, their fitness evaluation is still performed from an end-to-end communication--computation 
perspective. For each candidate policy, the resulting upload completion times are propagated as branch-entry release times to the downstream DAG, where a release-aware DAG 
scheduler is then executed. Therefore, the proposed lightweight methods are not conventional single-stage heuristics. Instead, they are DAG-aware approximations of joint scheduling that 
preserve the impact of communication decisions on downstream computation while significantly lowering the optimization cost.

\subsection{\texttt{GA-DACS}: Policy Representation}

In \texttt{GA-DACS}, each chromosome encodes only the policy parameters associated with the communication stage, i.e.,
\begin{equation}
\boldsymbol{\theta}_{\mathrm{DACS}}
=
\bm{\alpha}
\in\mathbb{R}^{8}.
\label{eq:dacs_policy_vector}
\end{equation}
Compared with the full joint policy in \eqref{eq:policy_vector}, the DAG-priority weights \(\bm{\beta}\) and the core-mapping weights \(\bm{\mu}\) are removed. Therefore, the search dimension is reduced from 20 to 8. This reduction directly decreases the GA search burden and makes \texttt{GA-DACS} suitable when the computation-side scheduler is required to remain simple, stable, or implementation-friendly.

Although the chromosome only parameterizes the communication scheduler, the communication decision is not optimized in isolation. The same normalized communication feature vector in \eqref{eq:comm_feature} is adopted, and each candidate stream is scored by
\begin{equation}
\Psi_k^{\mathrm{DACS}}(t,f)
=
\sum_{i=1}^{8}\alpha_i z_{k,i}(t,f).
\label{eq:dacs_comm_score}
\end{equation}
For each available RB \((t,f)\), the scheduled stream is selected as
\begin{equation}
k_{t,f}^{\star}
=
\arg\max_{k\in\mathcal{A}(t,f)}
\Psi_k^{\mathrm{DACS}}(t,f),
\label{eq:dacs_comm_decision}
\end{equation}
where \(\mathcal{A}(t,f)\) is the eligible stream set defined in \eqref{eq:eligible_set}. If \(\mathcal{A}(t,f)=\varnothing\), the corresponding RB remains idle. Otherwise, the selected stream updates its remaining payload according to the communication model. Repeating this process across all RBs within the scheduling horizon yields the upload completion time \(\tau_k\) of each stream.

The key difference from a purely communication-oriented heuristic is that the learned score in \eqref{eq:dacs_comm_score} is evaluated according to the final end-to-end performance, rather than the instantaneous transmission performance alone. Therefore, \texttt{GA-DACS} can implicitly learn whether it is more beneficial to prioritize large payloads, high-rate streams, long downstream branches, or synchronization-sensitive streams under the current system configuration.

\subsection{\texttt{GA-DACS}: Decoding}

After the communication-stage decoding, \texttt{GA-DACS} follows the same release-time propagation mechanism as \texttt{GA-Joint}, where the upload completion times are mapped to the release times 
of the corresponding branch-entry nodes according to \eqref{eq:node_release}. However, unlike \texttt{GA-Joint}, \texttt{GA-DACS} does not further optimize the ready-node priority rule or the 
core-mapping rule. Instead, it employs a deterministic release-aware greedy DAG scheduler to generate the computation schedule.

Specifically, the scheduler maintains a ready-node set and selects ready nodes according to a fixed deterministic topological priority rule. For each selected node \(v\), it evaluates all candidate cores and selects the one that yields the earliest feasible finish time:
\begin{equation}
a_v
=
\arg\min_{c\in\mathcal{C}}
\left(
\hat{s}_{v,c}+p_v
\right),
\label{eq:dacs_core_selection}
\end{equation}
where
\begin{equation}
\hat{s}_{v,c}
=
\max\left\{
\chi_c,\ \rho_v,\ 
\max_{u\in\mathrm{Pred}(v)}
\hat{\Phi}_{u\rightarrow v}^{(c)}
\right\}.
\label{eq:dacs_start_time}
\end{equation}

After the selected node is assigned to the chosen core, its start time, finish time, and the corresponding core availability are updated. The procedure continues until all DAG 
nodes are scheduled. In this way, \texttt{GA-DACS} preserves the release-time coupling between communication and computation, while avoiding the additional GA search over DAG ordering 
and core mapping. Since the individual completion time of each stream is propagated to its corresponding branch-entry node, streams that finish uploading earlier can activate their 
downstream DAG branches earlier, enabling partial communication--computation overlap at the schedule level.

\subsection{\texttt{GA-DAG}: Policy Representation}

In addition to \texttt{GA-DACS}, this paper further considers another lightweight variant \texttt{GA-DAG}, which reduces the search burden from the communication side. Compared with 
\texttt{GA-DACS}, \texttt{GA-DAG} adopts a fixed greedy communication scheduler and uses GA only to optimize the DAG-stage scheduling policy. Specifically, the chromosome of \texttt{GA-DAG} is defined as
\begin{equation}
\boldsymbol{\theta}_{\mathrm{DAG}}
=
\big[\bm{\beta},\bm{\mu}\big]
\in\mathbb{R}^{12}.
\label{eq:dag_policy_vector}
\end{equation}
Compared with the full joint policy in \eqref{eq:policy_vector}, the communication-stage weights \(\bm{\alpha}\) are removed. Therefore, the search dimension is reduced from 20 to 12.

The motivation of \texttt{GA-DAG} is complementary to that of \texttt{GA-DACS}. \texttt{GA-DACS} optimizes the communication scheduler while keeping the computation scheduler greedy, whereas \texttt{GA-DAG} keeps the communication scheduler fixed and optimizes the computation-side DAG execution 
policy. This design is useful when the communication scheduler is required to remain simple or when the main scheduling flexibility is expected to come from DAG-node ordering and core mapping.

\subsection{\texttt{GA-DAG}: Decoding}

In \texttt{GA-DAG}, the communication stage is generated by a fixed greedy scheduler. At each RB \((t,f)\), the eligible stream set \(\mathcal{A}(t,f)\) is first constructed according to \eqref{eq:eligible_set}. If \(\mathcal{A}(t,f)=\varnothing\), the corresponding RB remains idle. Otherwise, the scheduler selects the eligible stream with the largest remaining payload:
\begin{equation}
k_{t,f}^{\star}
=
\arg\max_{k\in\mathcal{A}(t,f)} Q_k(t,f).
\label{eq:gadag_comm_decision}
\end{equation}
After assigning stream \(k_{t,f}^{\star}\) to RB \((t,f)\), its remaining payload is updated according to \eqref{eq:queue_rb_update}--\eqref{eq:queue_next_slot}. Repeating this process across all RBs within the scheduling horizon yields the upload completion times \(\{\tau_k\}\). These completion times are then propagated to the corresponding branch-entry nodes as release times through \eqref{eq:node_release}.

After release-time propagation, \texttt{GA-DAG} performs DAG scheduling using the chromosome \(\boldsymbol{\theta}_{\mathrm{DAG}}\). For each ready node \(v\in\mathcal{R}\), the same DAG feature vector in \eqref{eq:dag_feature} is adopted. The ready-node priority score is defined as
\begin{equation}
\Psi_v^{\mathrm{DAG}}
=
\sum_{i=1}^{8}\beta_i q_{v,i}.
\label{eq:gadag_node_score}
\end{equation}
The next node to be scheduled is selected as
\begin{equation}
v^{\star}
=
\arg\max_{v\in\mathcal{R}}
\Psi_v^{\mathrm{DAG}}.
\label{eq:gadag_node_decision}
\end{equation}

After selecting \(v^{\star}\), the scheduler evaluates all candidate cores. If \(v^{\star}\) is tentatively assigned to core \(c\in\mathcal{C}\), its candidate start time is computed by \eqref{eq:candidate_start} and the corresponding candidate finish time is given by \eqref{eq:candidate_finish}.
To jointly account for latency, cross-core transfer, off-chip access, locality preservation, and core availability, the mapping score is defined as
\begin{equation}
\Psi_{v^{\star},c}^{\mathrm{DAG\text{-}map}}
=
\hat{f}_{v^{\star},c}
+\mu_1\kappa_{v^{\star},c}^{\mathrm{cross}}
+\mu_2\Delta_{v^{\star},c}^{\mathrm{off}}
-\mu_3\kappa_{v^{\star},c}^{\mathrm{same}}
+\mu_4\chi_c.
\label{eq:gadag_mapping_score}
\end{equation}
The selected core is then given by
\begin{equation}
a_{v^{\star}}
=
\arg\min_{c\in\mathcal{C}}
\Psi_{v^{\star},c}^{\mathrm{DAG\text{-}map}}.
\label{eq:gadag_core_decision}
\end{equation}
After the node is assigned, its start time, finish time, and the corresponding core availability are updated. This process continues until all DAG nodes are completed.

\subsection{Fitness Evaluation of Lightweight Variants}

For \texttt{GA-DACS}, each chromosome \(\boldsymbol{\theta}_{\mathrm{DACS}}\) first generates a communication schedule and obtains the stream completion times \(\{\tau_k\}\). These completion times are then used as DAG release times, and the fixed release-aware DAG scheduler produces the final computation schedule. For \texttt{GA-DAG}, the communication schedule is first generated by the fixed greedy communication scheduler, and the chromosome 
\(\boldsymbol{\theta}_{\mathrm{DAG}}\) is then used to determine the DAG ready-node priority and core-mapping decisions.

Both lightweight variants are evaluated using the same end-to-end objective structure:
\begin{equation}
\begin{aligned}
F_{\mathrm{DACS}}(\boldsymbol{\theta}_{\mathrm{DACS}})
&=
T_{\mathrm{e2e}}(\boldsymbol{\theta}_{\mathrm{DACS}})
+
\lambda P_{\mathrm{sync}}(\boldsymbol{\theta}_{\mathrm{DACS}}) \\
&\quad
+
P_{\mathrm{inv}}(\boldsymbol{\theta}_{\mathrm{DACS}}),
\end{aligned}
\label{eq:dacs_fitness}
\end{equation}
and
\begin{equation}
\begin{aligned}
F_{\mathrm{DAG}}(\boldsymbol{\theta}_{\mathrm{DAG}})
&=
T_{\mathrm{e2e}}(\boldsymbol{\theta}_{\mathrm{DAG}})
+
\lambda P_{\mathrm{sync}}(\boldsymbol{\theta}_{\mathrm{DAG}})\\
&\quad
+
P_{\mathrm{inv}}(\boldsymbol{\theta}_{\mathrm{DAG}}).
\end{aligned}
\label{eq:gadag_fitness}
\end{equation}
Here, \(P_{\mathrm{inv}}(\cdot)\) is applied only when not all streams finish uploading within the scheduling horizon.

This fitness design is important because it allows the reduced chromosome to be judged from a computation-aware perspective. For \texttt{GA-DACS}, a communication policy that achieves high instantaneous throughput may still be penalized if it delays critical branches, increases synchronization mismatch, or postpones the release of downstream fusion nodes. Conversely, a policy with slightly lower communication efficiency may obtain a better fitness value if it enables earlier DAG execution and reduces the final end-to-end latency. For \texttt{GA-DAG}, the fixed communication schedule determines the branch release times, while 
the GA-optimized DAG policy improves downstream execution by adapting ready-node ordering and core mapping to these release times.

\begin{algorithm}[!t]
\caption{Lightweight \texttt{GA-DACS} Scheduling}
\label{alg:ga_dacs}
\begin{algorithmic}[1]
\Require $G=(\mathcal{V},\mathcal{E})$, $\mathcal{K}$, $\mathcal{F}$, $\mathcal{C}$, 
$\{D_k\}_{k\in\mathcal{K}}$, $\{\gamma_k(t,f)\}$, $T_{\max}$, $\lambda$, 
$P_{\mathrm{DACS}}$, $G_{\max}^{\mathrm{DACS}}$
\Ensure $\boldsymbol{\theta}_{\mathrm{DACS}}^{\star},\ J_{\mathrm{DACS}}^{\star}$

\State Initialize population
$\mathcal{P}^{(0)}=\{\boldsymbol{\theta}_{\mathrm{DACS}}^{(p,0)}\}_{p=1}^{P_{\mathrm{DACS}}}$,
where $\boldsymbol{\theta}_{\mathrm{DACS}}^{(p,0)}=\bm{\alpha}^{(p,0)}$.
\label{ln:dacs:init-pop}

\For{$g=0$ to $G_{\max}^{\mathrm{DACS}}-1$}
    \For{$p=1$ to $P_{\mathrm{DACS}}$}

        \State Execute Lines~\ref{ln:joint:eval-init}--\ref{ln:joint:comm-end} of 
        Algorithm~\ref{alg:ga_joint}, with 
        $\boldsymbol{\theta}^{(p,g)}$ replaced by 
        $\boldsymbol{\theta}_{\mathrm{DACS}}^{(p,g)}=\bm{\alpha}^{(p,g)}$, and
        $\Psi_k^{\mathrm{comm}}$ replaced by $\Psi_k^{\mathrm{DACS}}$.
        \label{ln:dacs:comm}

        \If{$\exists k\in\mathcal{K}: \tau_k^{(p,g)}=+\infty$}
            \State $F_{\mathrm{DACS}}(\boldsymbol{\theta}_{\mathrm{DACS}}^{(p,g)})
            \gets P_{\mathrm{inv}}$.
            \State \textbf{continue}
        \EndIf
        \label{ln:dacs:infeas}

        \State Propagate $\{\tau_k^{(p,g)}\}$ to branch-entry release times 
        $\{\rho_v^{(p,g)}\}$ according to \eqref{eq:node_release}.
        \label{ln:dacs:release}

        \State $\chi_c\gets 0,\ \forall c\in\mathcal{C}$; \quad 
        $\mathcal{U}\gets\mathcal{V}$.
        \label{ln:dacs:dag-init}

        \While{$\mathcal{U}\neq\varnothing$}
            \State $\mathcal{R}\gets
            \{v\in\mathcal{U}\mid \mathrm{Pred}(v)\cap\mathcal{U}=\varnothing\}$.

            \State Select $v^\star\in\mathcal{R}$ according to a fixed deterministic 
            topological priority rule.
            \label{ln:dacs:node-select}

            \State Select the core with the earliest feasible finish time:
            \[
            c^\star\gets
            \arg\min_{c\in\mathcal{C}}
            \left(\hat{s}_{v^\star,c}+p_{v^\star}\right),
            \]
            where $\hat{s}_{v^\star,c}$ is computed by \eqref{eq:dacs_start_time}.
            \label{ln:dacs:core-select}

            \State $a_{v^\star}\gets c^\star$; 
            $s_{v^\star}\gets\hat{s}_{v^\star,c^\star}$; 
            $f_{v^\star}\gets s_{v^\star}+p_{v^\star}$.

            \State $\chi_{c^\star}\gets f_{v^\star}$; \quad
            $\mathcal{U}\gets\mathcal{U}\setminus\{v^\star\}$.
        \EndWhile
        \label{ln:dacs:dag-end}

        \State Compute $T_{\mathrm{e2e}}^{(p,g)}$ and $P_{\mathrm{sync}}^{(p,g)}$ as in 
        Lines~\ref{ln:joint:obj-begin}--\ref{ln:joint:obj-end} of 
        Algorithm~\ref{alg:ga_joint}.

        \State $F_{\mathrm{DACS}}(\boldsymbol{\theta}_{\mathrm{DACS}}^{(p,g)})
        \gets
        T_{\mathrm{e2e}}^{(p,g)}
        +\lambda P_{\mathrm{sync}}^{(p,g)}$.
        \label{ln:dacs:fitness}

    \EndFor

    \State Execute Lines~\ref{ln:joint:ga-begin}--\ref{ln:joint:ga-end} of 
    Algorithm~\ref{alg:ga_joint}, with $F$ replaced by $F_{\mathrm{DACS}}$.
    \label{ln:dacs:ga-update}

\EndFor

\State $\boldsymbol{\theta}_{\mathrm{DACS}}^{\star}\gets
\arg\min\limits_{\boldsymbol{\theta}_{\mathrm{DACS}}\in\cup_g\mathcal{P}^{(g)}}
F_{\mathrm{DACS}}(\boldsymbol{\theta}_{\mathrm{DACS}})$.

\State $J_{\mathrm{DACS}}^{\star}\gets
F_{\mathrm{DACS}}(\boldsymbol{\theta}_{\mathrm{DACS}}^{\star})$.

\State \Return $\boldsymbol{\theta}_{\mathrm{DACS}}^{\star},\ J_{\mathrm{DACS}}^{\star}$.

\end{algorithmic}
\end{algorithm}

\subsection{Optimization Procedure}

The optimization process of both lightweight variants follows the same population-based policy-search principle as \texttt{GA-Joint}, but with smaller chromosomes and simplified decoders. For \texttt{GA-DACS}, each chromosome contains only the 8 communication weights in \eqref{eq:dacs_policy_vector}. During fitness evaluation, the communication schedule is decoded using \eqref{eq:dacs_comm_score} and \eqref{eq:dacs_comm_decision}, and the resulting 
stream completion times are passed to the fixed release-aware DAG scheduler. The GA updates the population according to the obtained fitness values through selection, crossover, mutation, and elite preservation. The overall procedure is summarized in Algorithm~\ref{alg:ga_dacs}. For \texttt{GA-DAG}, each chromosome contains the 12 DAG-stage weights in \eqref{eq:dag_policy_vector}. During fitness evaluation, the communication schedule is first generated by the fixed greedy communication scheduler in \eqref{eq:gadag_comm_decision}. The resulting release times are then propagated to the DAG, and the DAG execution schedule is decoded according to \eqref{eq:gadag_node_score}--\eqref{eq:gadag_core_decision}. The corresponding procedure is summarized in Algorithm~\ref{alg:ga_dag}.

\begin{algorithm}[!t]
\caption{Lightweight \texttt{GA-DAG} Scheduling}
\label{alg:ga_dag}
\begin{algorithmic}[1]
\Require $G=(\mathcal{V},\mathcal{E})$, $\mathcal{K}$, $\mathcal{F}$, $\mathcal{C}$, 
$\{D_k\}_{k\in\mathcal{K}}$, $\{\gamma_k(t,f)\}$, $T_{\max}$, $\lambda$, 
$P_{\mathrm{DAG}}$, $G_{\max}^{\mathrm{DAG}}$
\Ensure $\boldsymbol{\theta}_{\mathrm{DAG}}^{\star},\ J_{\mathrm{DAG}}^{\star}$

\State Initialize population
$\mathcal{P}^{(0)}=\{\boldsymbol{\theta}_{\mathrm{DAG}}^{(p,0)}\}_{p=1}^{P_{\mathrm{DAG}}}$,
where $\boldsymbol{\theta}_{\mathrm{DAG}}^{(p,0)}
=[\bm{\beta}^{(p,0)},\bm{\mu}^{(p,0)}]$.
\label{ln:gadag:init-pop}

\For{$g=0$ to $G_{\max}^{\mathrm{DAG}}-1$}
    \For{$p=1$ to $P_{\mathrm{DAG}}$}

        \State Generate the OFDMA upload schedule by scanning all RBs and applying the 
        fixed greedy rule in \eqref{eq:gadag_comm_decision} over 
        \(\mathcal{A}(t,f)\); update the payloads according to \eqref{eq:queue_rb_update}--\eqref{eq:queue_next_slot} and 
        obtain \(\{\tau_k^{(p,g)}\}\).
        \label{ln:gadag:comm}

        \If{$\exists k\in\mathcal{K}: \tau_k^{(p,g)}=+\infty$}
            \State $F_{\mathrm{DAG}}(\boldsymbol{\theta}_{\mathrm{DAG}}^{(p,g)})
            \gets P_{\mathrm{inv}}$.
            \State \textbf{continue}
        \EndIf
        \label{ln:gadag:infeas}

        \State Propagate $\{\tau_k^{(p,g)}\}$ to branch-entry release times 
        $\{\rho_v^{(p,g)}\}$ according to \eqref{eq:node_release}.
        \label{ln:gadag:release}

        \State $\chi_c\gets 0,\ \forall c\in\mathcal{C}$; \quad
        $\mathcal{U}\gets\mathcal{V}$.
        \label{ln:gadag:dag-init}

        \While{$\mathcal{U}\neq\varnothing$}
            \State $\mathcal{R}\gets
            \{v\in\mathcal{U}\mid \mathrm{Pred}(v)\cap\mathcal{U}=\varnothing\}$.

            \State $v^\star\gets
            \arg\max\limits_{v\in\mathcal{R}}
            \Psi_v^{\mathrm{DAG}}(\boldsymbol{\theta}_{\mathrm{DAG}}^{(p,g)})$.
            \label{ln:gadag:node-select}

            \State $c^\star\gets
            \arg\min\limits_{c\in\mathcal{C}}
            \Psi_{v^\star,c}^{\mathrm{DAG\text{-}map}}
            (\boldsymbol{\theta}_{\mathrm{DAG}}^{(p,g)})$.
            \label{ln:gadag:core-select}

            \State $a_{v^\star}\gets c^\star$.

            \State $s_{v^\star}\gets
            \max\Bigl\{
            \rho_{v^\star}^{(p,g)},\,
            \chi_{c^\star},\,
            \max\limits_{u\in\mathrm{Pred}(v^\star)}
            \hat{\Phi}_{u\to v^\star}^{(c^\star)}
            \Bigr\}$.
            \label{ln:gadag:start}

            \State $f_{v^\star}\gets s_{v^\star}+p_{v^\star}$.

            \State $\chi_{c^\star}\gets f_{v^\star}$; \quad
            $\mathcal{U}\gets\mathcal{U}\setminus\{v^\star\}$.
        \EndWhile
        \label{ln:gadag:dag-end}

        \State Compute $T_{\mathrm{e2e}}^{(p,g)}$ and $P_{\mathrm{sync}}^{(p,g)}$ as in 
        Lines~\ref{ln:joint:obj-begin}--\ref{ln:joint:obj-end} of 
        Algorithm~\ref{alg:ga_joint}.

        \State $F_{\mathrm{DAG}}(\boldsymbol{\theta}_{\mathrm{DAG}}^{(p,g)})
        \gets
        T_{\mathrm{e2e}}^{(p,g)}
        +\lambda P_{\mathrm{sync}}^{(p,g)}$.
        \label{ln:gadag:fitness}

    \EndFor

    \State Execute Lines~\ref{ln:joint:ga-begin}--\ref{ln:joint:ga-end} of 
    Algorithm~\ref{alg:ga_joint}, with $F$ replaced by $F_{\mathrm{DAG}}$.
    \label{ln:gadag:ga-update}

\EndFor

\State $\boldsymbol{\theta}_{\mathrm{DAG}}^{\star}\gets
\arg\min\limits_{\boldsymbol{\theta}_{\mathrm{DAG}}\in\cup_g\mathcal{P}^{(g)}}
F_{\mathrm{DAG}}(\boldsymbol{\theta}_{\mathrm{DAG}})$.

\State $J_{\mathrm{DAG}}^{\star}\gets
F_{\mathrm{DAG}}(\boldsymbol{\theta}_{\mathrm{DAG}}^{\star})$.

\State \Return $\boldsymbol{\theta}_{\mathrm{DAG}}^{\star},\ J_{\mathrm{DAG}}^{\star}$.

\end{algorithmic}
\end{algorithm}

\FloatBarrier

\subsection{Complexity Analysis of Lightweight Variants}

The main computational saving of the lightweight GA-based scheduling variants comes from the reduced chromosome dimension and the removal of part of the GA-optimized scheduling policy. For \texttt{GA-DACS}, the chromosome contains only the communication-stage weights \(\bm{\alpha}\in\mathbb{R}^{8}\). For \texttt{GA-DAG}, the chromosome contains only the DAG-stage weights \([\bm{\beta},\bm{\mu}]\in\mathbb{R}^{12}\). Both are smaller than the 20-dimensional chromosome used by \texttt{GA-Joint}.

It is worth noting that the greedy components in the two lightweight variants are also included in the decoding complexity. For \texttt{GA-DACS}, although the DAG execution is no longer optimized by GA, the release-aware greedy DAG scheduler still needs to repeatedly identify ready nodes and evaluate feasible core assignments, which contributes the \(N^2+CE\) term. For \texttt{GA-DAG}, the communication schedule is generated by a fixed greedy rule, but it still scans all RB resources and candidate streams, leading to the \(TFK\) term. Therefore, the lightweight variants reduce the GA search dimension and the number of tunable scheduling weights, while their per-fitness decoding order remains comparable to that of \texttt{GA-Joint}).

Accordingly, the total optimization complexity of \texttt{GA-DACS} is
\begin{equation}
\mathcal{O}\Big(P_{\mathrm{DACS}}G_{\max}^{\mathrm{DACS}}
\big(TFK+N^2+CE\big)\Big),
\label{eq:dacs_total_complexity}
\end{equation}
where \(P_{\mathrm{DACS}}\) and \(G_{\max}^{\mathrm{DACS}}\) denote the population size and maximum number of generations used by \texttt{GA-DACS}. Similarly, the total optimization complexity of \texttt{GA-DAG} is
\begin{equation}
\mathcal{O}\Big(P_{\mathrm{DAG}}G_{\max}^{\mathrm{DAG}}
\big(TFK+N^2+CE\big)\Big),
\label{eq:gadag_total_complexity}
\end{equation}
where \(P_{\mathrm{DAG}}\) and \(G_{\max}^{\mathrm{DAG}}\) denote the corresponding GA parameters of \texttt{GA-DAG}.

Compared with \texttt{GA-Joint}, the asymptotic decoding order of the lightweight variants remains the same because each fitness evaluation still requires both communication scheduling and DAG execution scheduling. However, their practical complexity is reduced through smaller chromosomes, fewer score components, and lower-dimensional GA search spaces. In addition, the simplified greedy decoders can reduce the constant factors in fitness evaluation, and the reduced search dimension allows smaller populations or fewer generations to be used in practice.

\section{Simulation Results}\label{sec:simulation_results}
This section evaluates the proposed GA-based scheduling algorithms with respect to end-to-end latency reduction. Besides the proposed algorithms, two baseline schemes are also considered: \texttt{Decoupled-Greedy} and \texttt{Joint-Greedy}. \texttt{Decoupled-Greedy} is a two-stage greedy baseline in which uplink communication and DAG execution on the multi-core accelerator are scheduled independently. In the communication stage, each available RB is assigned to the feasible sensing stream with the largest remaining payload, while DAG execution starts only after all streams have completed their uploads. This scheme represents a conventional decoupled design without explicit communication–computation overlap. \texttt{Joint-Greedy} is a release-aware greedy baseline without GA optimization. For each RB allocation, the scheduler tentatively assigns the current RB to each feasible sensing stream and estimates the resulting end-to-end latency via downstream release-aware DAG scheduling. The RB is then allocated to the stream that achieves the best one-step objective, thereby capturing communication–computation coupling in a local and myopic manner. The key distinctions of these schemes are summarized in Table~\ref{tab:scheme_comparison}.

\subsection{Illustrative Example}
In the simulations, the numbers of sensing streams and computing cores are set to \(K=6\) and \(C=4\), respectively. The branch lengths of the six stream-specific DAGs are specified as \([5,8,5,6,5,7]\). The streams are divided into two synchronization groups, namely, \(\{1,2,3\}\) and \(\{4,5,6\}\). Following the group-level alignment nodes, a fusion head composed of fusion, classifier, and output nodes is appended to complete the inference DAG. The input payload sizes are set to \([0.2,1.5,5,2,7,10]~\mathrm{kB}\). The scheduling horizon is set to \(T_{\max}=1000\) slots, where each slot consists of \(F=4\) orthogonal subcarriers, each with bandwidth \(180~\mathrm{kHz}\). The SINR of each stream is randomly generated over \([5,20]\) dB, with the SINR threshold set to \(6\) dB. The achievable rate is evaluated using a Shannon-type model with unit gap, and the spectral efficiency is capped at \(8~\mathrm{bit/s/Hz}\). For DAG execution, the computation time of each node is randomly generated over \([1,11]\) ms. The on-chip and off-chip read/write delays are generated over \([0.1,0.6]\) ms and \([2,8]\) ms, respectively. The synchronization penalty coefficient is set to \(\lambda=0.05\). For all GA-based schemes, the population size, maximum number of generations, and weight search range are set to \(40\), \(35\), and \([-10,10]\), respectively.

\begin{figure}
\centering
\includegraphics[width=0.9\linewidth]{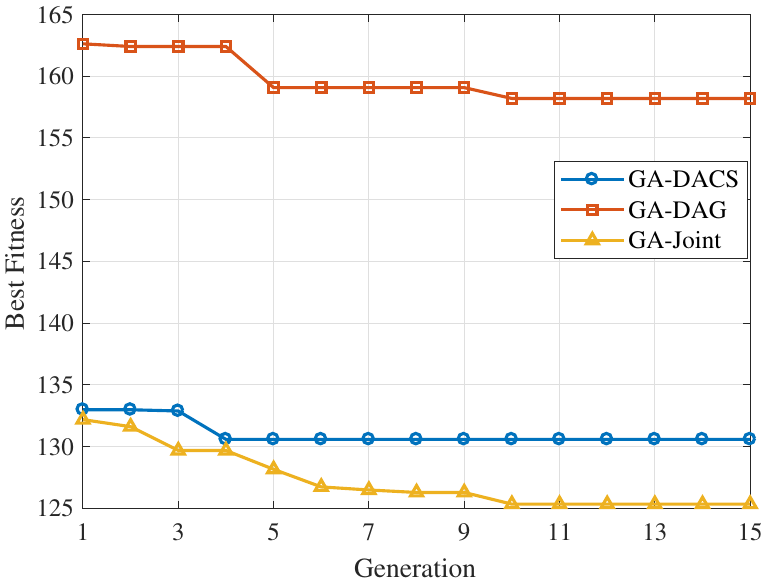}
\caption{Convergence behavior of the three GA-based scheduling schemes.}\label{Fig2}
\end{figure}

Fig.~\ref{Fig2} illustrates the convergence behavior of the three GA-based schemes, namely, \texttt{GA-DAG}, \texttt{GA-DACS}, and \texttt{GA-Joint}. The vertical axis denotes the best fitness value achieved up to each generation, where a smaller value indicates lower end-to-end latency and synchronization penalty. All three schemes exhibit monotonic improvement during the evolutionary process, demonstrating the effectiveness of the adopted chromosome representations and fitness evaluation for the considered joint scheduling problem. Among them, \texttt{GA-DACS} converges rapidly due to its reduced search space associated with communication-policy optimization, whereas \texttt{GA-DAG} optimizes computation-side priority and core assignment under fixed release times. In contrast, \texttt{GA-Joint} explores the largest search space by jointly adapting communication scheduling, DAG priority, and core mapping. Nevertheless, it achieves the lowest final fitness value, indicating that full cross-stage optimization provides the highest flexibility in exploiting communication--computation overlap.

\begin{figure*}[!t]
\centering
\includegraphics[width=\textwidth]{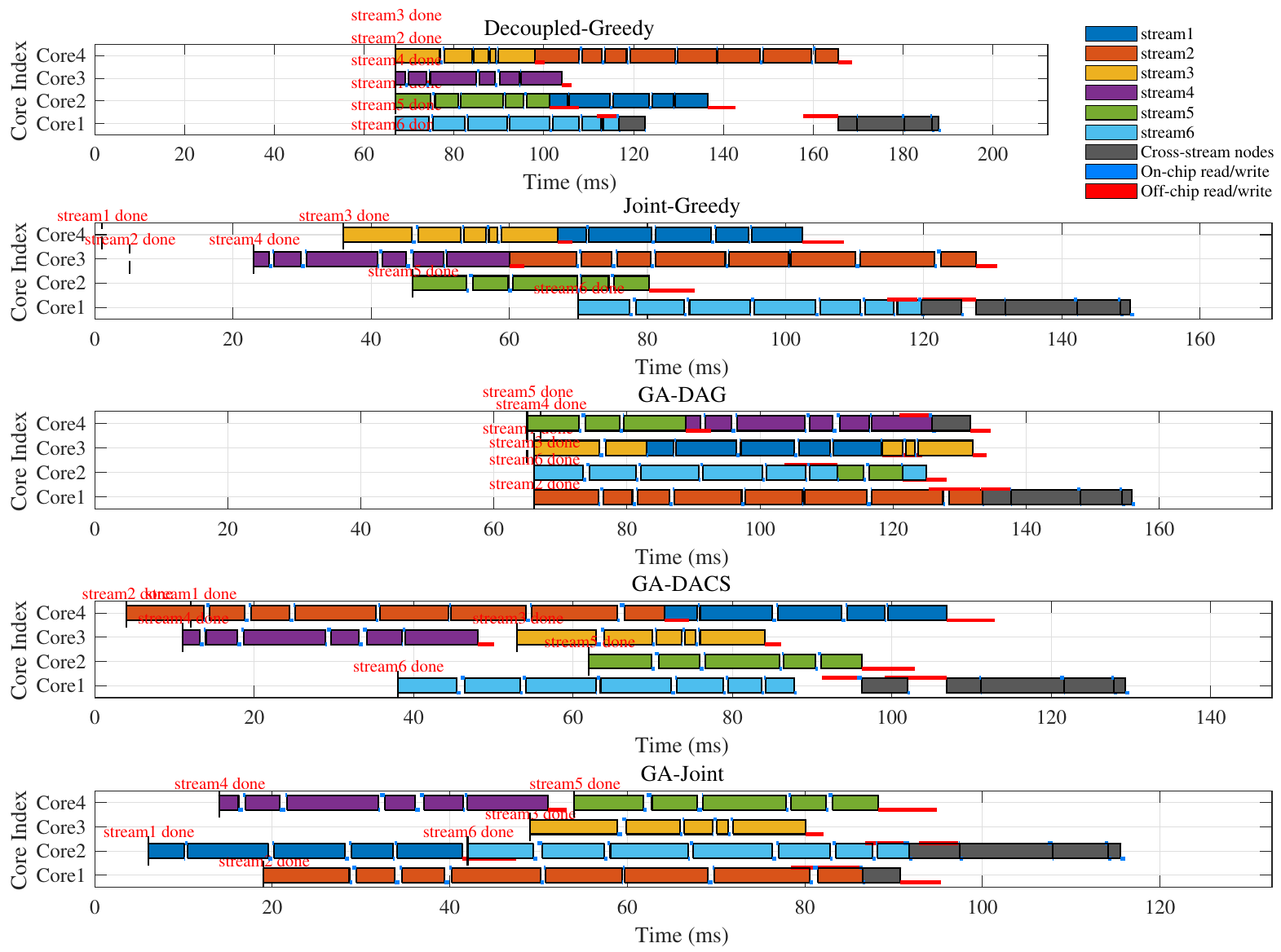}
\caption{Comparison of end-to-end execution timelines under the evaluated scheduling schemes.}
\label{Fig3}
\end{figure*}

Fig.~\ref{Fig3} compares the end-to-end execution timelines of the five scheduling schemes. Compared with \texttt{Decoupled-Greedy}, all the other schemes are release-aware and activate different DAG branches according to their respective stream completion times, thereby enabling partial communication--computation overlap. Although \texttt{Joint-Greedy} accounts for such coupling through local one-step estimation, its myopic decision rule limits the overall scheduling efficiency. The GA-based schemes generate more compact execution schedules. Specifically, \texttt{GA-DAG} improves computation-side task ordering and core mapping under fixed communication decisions, whereas \texttt{GA-DACS} optimizes the communication policy from a DAG-aware perspective and advances the release of computation-critical branches. Among all schemes, \texttt{GA-Joint} achieves the most compact schedule by jointly optimizing communication scheduling, DAG priority, and core assignment. The resulting reduction in end-to-end latency confirms the benefit of full joint scheduling.


\begin{table*}[t]
\centering
\caption{Comparison of the evaluated scheduling schemes}
\label{tab:scheme_comparison}
\renewcommand{\arraystretch}{1.15}
\setlength{\tabcolsep}{6pt}
\scriptsize
\begin{tabular}{l|p{4.55cm}|p{4.6cm}|p{4.0cm}}
\hline
\textbf{Scheme} 
& \textbf{Communication Scheduling} 
& \textbf{DAG Execution} 
& \textbf{GA-Optimized Components} \\
\hline
\textbf{\texttt{Joint-Greedy}} 
& DAG-aware greedy 
& Release-aware greedy execution 
& None  \\
\hline
\textbf{\texttt{Decoupled-Greedy}} 
& Remaining-payload-based greedy 
& Waits until all streams finish uploading 
& None  \\
\hline
\textbf{\texttt{GA-DACS}} 
& GA-optimized communication scoring policy 
& Release-aware greedy execution 
& Communication scoring weights \\
\hline
\textbf{\texttt{GA-DAG}} 
& Fixed greedy communication scheduling 
& GA-optimized DAG priority and core mapping 
& DAG priority and core mapping weights \\
\hline
\textbf{\texttt{GA-Joint}} 
& GA-optimized communication scoring policy 
& GA-optimized DAG priority and core mapping 
& Communication scoring, DAG priority, and core mapping weights  \\
\hline
\end{tabular}
\end{table*}

\subsection{Impact of System-Scaling Factors}
\begin{figure}[!t]
\centering
\includegraphics[width=0.9\linewidth]{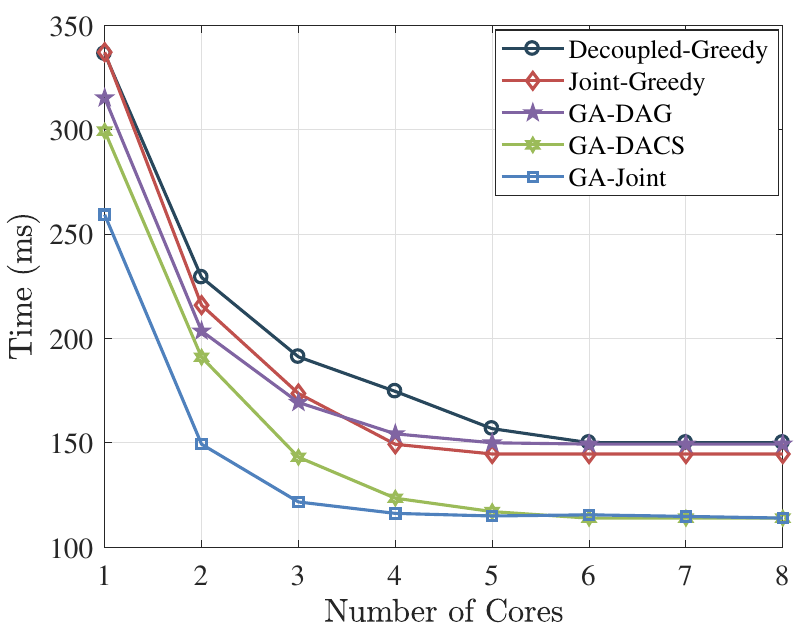}
\caption{Impact of the number of accelerator cores on end-to-end latency.}
\label{Fig4}
\end{figure}

\begin{figure}[!t]
\centering
\includegraphics[width=0.9\linewidth]{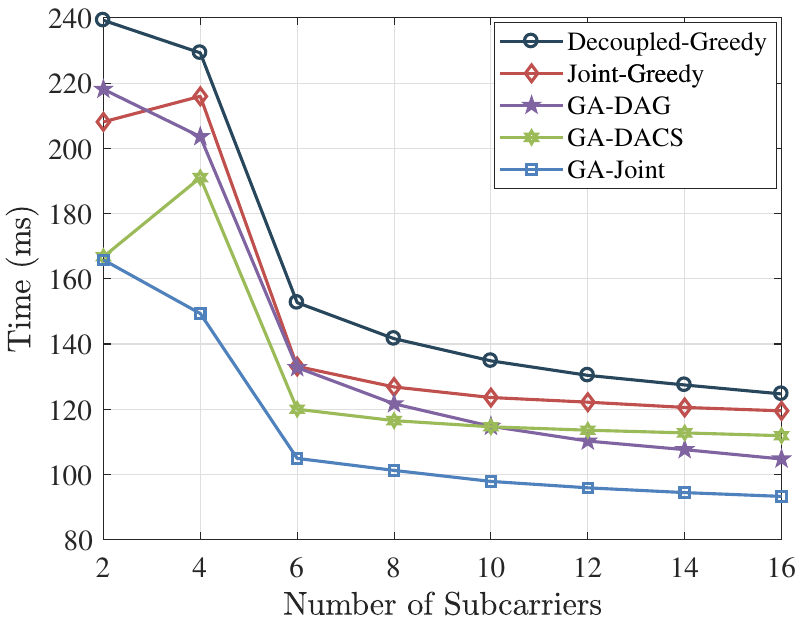}
\caption{Impact of the number of subcarriers on end-to-end latency.}
\label{Fig5}
\end{figure}

\begin{figure}[!t]
\centering
\includegraphics[width=0.9\linewidth]{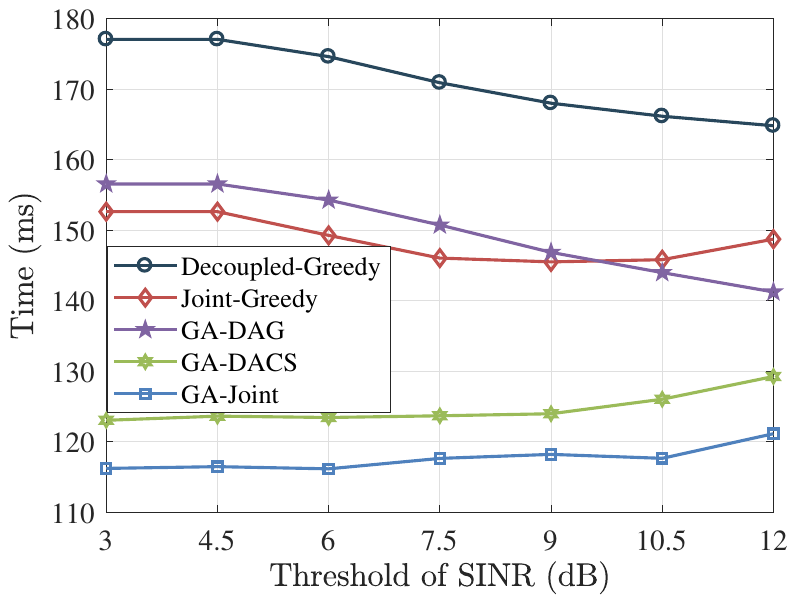}
\caption{Impact of the SINR threshold on end-to-end latency.}
\label{Fig6}
\end{figure}

Figs.~\ref{Fig4}, \ref{Fig5}, and \ref{Fig6} illustrate the impact of three scaling dimensions on end-to-end latency, namely, the number of accelerator cores, the number of subcarriers, and the SINR threshold, respectively. Several trends can be observed. First, increasing the number of accelerator cores reduces the latency of all schemes, since more branch and fusion operations can be executed in parallel. This gain, however, gradually saturates once the communication stage becomes the dominant bottleneck. Second, increasing the number of subcarriers improves the uplink service capability and shortens stream completion times, thereby reducing both the communication delay and the release times of downstream DAG nodes. Third, the impact of the SINR threshold is not necessarily monotonic, since a higher threshold may reduce RB utilization by imposing stricter feasibility constraints, while also improving allocation quality by excluding poor transmission opportunities.

The proposed GA-based schemes consistently retain their advantages across the considered scaling regimes. \texttt{Decoupled-Greedy} exhibits the largest latency, as it cannot exploit release-time diversity across different streams. \texttt{Joint-Greedy} improves upon this baseline by enabling release-aware execution, but its performance gain is constrained by locally myopic communication decisions. \texttt{GA-DAG} and \texttt{GA-DACS} provide complementary benefits: the former improves accelerator-side scheduling under fixed stream release times, whereas the latter reshapes stream release times through DAG-aware communication scheduling. In most cases, \texttt{GA-Joint} achieves the lowest latency by jointly coordinating communication scheduling, DAG priority, and core assignment. These results demonstrate the robustness of the proposed joint design under varying communication and computation resources. Note that although \texttt{GA-Joint} has a more expressive policy space, its larger search dimension may make finite-generation GA optimization more difficult. Therefore, \texttt{GA-DACS} can occasionally obtain slightly lower latency under the same GA budget due to its reduced search space and more stable convergence.

\subsection{Impact of Communication Loads}
Table~\ref{tab:latency_data_core} reports the impact of varying communication loads on end-to-end latency. The results are consistent with the preceding discussion and lead to three observations. First, as payload size increases and across both homogeneous and heterogeneous data settings, the proposed schemes maintain clear advantages. Second, increasing the number of accelerator cores generally reduces latency across all schemes, but the marginal benefit diminishes once communication dominates the end-to-end delay. Third, among the five schemes, \texttt{GA-Joint} achieves the lowest latency in most tested settings, while \texttt{GA-DACS} remains highly competitive and sometimes approaches the performance of full joint optimization. This suggests that communication-side DAG awareness, together with the GA-based optimization strategy, can capture a substantial fraction of the achievable latency reduction. Table~\ref{tab:latency_nodes_core} further reports the end-to-end latency under different branch-length configurations. Compared with Table~\ref{tab:latency_data_core}, this set of results mainly highlights the impact of computational heterogeneity on cross-stage scheduling. The results show that the proposed \texttt{GA-Joint} still achieves a clear advantage, while the lightweight \texttt{GA-DAG} and \texttt{GA-DACS} also perform well in most cases.

\begin{table*}[t]
\centering
\caption{End-to-end latency under different communication loads.}
\label{tab:latency_data_core}

\setlength{\tabcolsep}{0.8pt}
\renewcommand{\arraystretch}{1.08}
\scriptsize

\begin{tabular}{@{}
>{\raggedright\arraybackslash}p{0.185\textwidth}
@{\hspace{2pt}}
cc
>{\centering\arraybackslash}p{0.042\textwidth}
>{\centering\arraybackslash}p{0.047\textwidth}
>{\centering\arraybackslash}p{0.047\textwidth}
@{\hspace{4pt}}
cc
>{\centering\arraybackslash}p{0.042\textwidth}
>{\centering\arraybackslash}p{0.047\textwidth}
>{\centering\arraybackslash}p{0.047\textwidth}
@{\hspace{4pt}}
cc
>{\centering\arraybackslash}p{0.042\textwidth}
>{\centering\arraybackslash}p{0.047\textwidth}
>{\centering\arraybackslash}p{0.047\textwidth}
@{}}
\toprule
\multicolumn{1}{c}{\multirow{2}{*}{Payload configuration}}
& \multicolumn{5}{c}{$C=2$}
& \multicolumn{5}{c}{$C=4$}
& \multicolumn{5}{c}{$C=6$} \\
\cmidrule(lr){2-6}\cmidrule(lr){7-11}\cmidrule(lr){12-16}
& \textbf{\makecell[c]{Decoupled-\\Greedy}}
& \textbf{\makecell[c]{Joint-\\Greedy}}
& \textbf{\makecell[c]{GA-\\DAG}}
& \textbf{\makecell[c]{GA-\\DACS}}
& \textbf{\makecell[c]{GA-\\Joint}}
& \textbf{\makecell[c]{Decoupled-\\Greedy}}
& \textbf{\makecell[c]{Joint-\\Greedy}}
& \textbf{\makecell[c]{GA-\\DAG}}
& \textbf{\makecell[c]{GA-\\DACS}}
& \textbf{\makecell[c]{GA-\\Joint}}
& \textbf{\makecell[c]{Decoupled-\\Greedy}}
& \textbf{\makecell[c]{Joint-\\Greedy}}
& \textbf{\makecell[c]{GA-\\DAG}}
& \textbf{\makecell[c]{GA-\\DACS}}
& \textbf{\makecell[c]{GA-\\Joint}} \\
\midrule

\([0.2,1.5,5,2,7,10]\times5\) kB
& 493.86 & 438.13 & 472.90 & 365.68 & 328.10
& 439.26 & 402.73 & 418.85 & 318.07 & 316.85
& 412.76 & 402.33 & 412.16 & 316.93 & 315.41 \\

\([0.2,1.5,5,2,7,10]\) kB
& 229.25 & 215.94 & 203.56 & 191.00 & 149.33
& 174.63 & 149.29 & 154.3 & 123.47 & 116.19 
& 150.05 & 144.62 & 149.38 & 113.90 & 115.51 \\

\([0.2,1.5,5,2,7,10]/2\) kB
& 191.05 & 192.86 & 166.83 & 170.59 & 142.24
& 140.06 & 129.24 & 119.45 & 115.03 & 97.331
& 113.56 & 110.55 & 113.16 & 90.65 & 90.56 \\

\([4,4,4,4,4,4]\) kB
& 219.85 & 223.30 & 194.96 & 174.68 & 158.25
& 168.86 & 153.34 & 148.05 & 124.62 & 119.61
& 142.36 & 137.67 & 141.03 & 109.14 & 109.14 \\

\([0.2,\cancel{1.5},5,\cancel{2.0},7,\cancel{10}]\) kB
& 117.63 & 111.43 & 102.40 & 99.634 & 84.198
& 88.879 & 89.505 & 88.201 & 80.769 & 81.369
& 88.879 & 89.505 & 80.769 & 88.201 & 80.588 \\

\bottomrule
\end{tabular}
\end{table*}

\begin{table*}[t]
\centering
\caption{End-to-end latency under different branch-length configurations.}
\label{tab:latency_nodes_core}

\setlength{\tabcolsep}{0.8pt}
\renewcommand{\arraystretch}{1.08}
\scriptsize

\begin{tabular}{@{}
>{\raggedright\arraybackslash}p{0.185\textwidth}
@{\hspace{2pt}}
cc
>{\centering\arraybackslash}p{0.042\textwidth}
>{\centering\arraybackslash}p{0.047\textwidth}
>{\centering\arraybackslash}p{0.047\textwidth}
@{\hspace{4pt}}
cc
>{\centering\arraybackslash}p{0.042\textwidth}
>{\centering\arraybackslash}p{0.047\textwidth}
>{\centering\arraybackslash}p{0.047\textwidth}
@{\hspace{4pt}}
cc
>{\centering\arraybackslash}p{0.042\textwidth}
>{\centering\arraybackslash}p{0.047\textwidth}
>{\centering\arraybackslash}p{0.047\textwidth}
@{}}
\toprule
\multicolumn{1}{c}{\multirow{2}{*}{Stream-specific branch}}
& \multicolumn{5}{c}{$C=2$}
& \multicolumn{5}{c}{$C=4$}
& \multicolumn{5}{c}{$C=6$} \\
\cmidrule(lr){2-6}\cmidrule(lr){7-11}\cmidrule(lr){12-16}
& \textbf{\makecell[c]{Decoupled-\\Greedy}}
& \textbf{\makecell[c]{Joint-\\Greedy}}
& \textbf{\makecell[c]{GA-\\DAG}}
& \textbf{\makecell[c]{GA-\\DACS}}
& \textbf{\makecell[c]{GA-\\Joint}}
& \textbf{\makecell[c]{Decoupled-\\Greedy}}
& \textbf{\makecell[c]{Joint-\\Greedy}}
& \textbf{\makecell[c]{GA-\\DAG}}
& \textbf{\makecell[c]{GA-\\DACS}}
& \textbf{\makecell[c]{GA-\\Joint}}
& \textbf{\makecell[c]{Decoupled-\\Greedy}}
& \textbf{\makecell[c]{Joint-\\Greedy}}
& \textbf{\makecell[c]{GA-\\DAG}}
& \textbf{\makecell[c]{GA-\\DACS}}
& \textbf{\makecell[c]{GA-\\Joint}} \\
\midrule

\([5,\,5,\,5,\,5,\,5,\,5]\)
& 199.41 & 199.5 & 174.19 & 180.24 & 125.18
& 156.77 & 128.77 & 140.65  & 112.22 & 110.6
& 132.2 & 122.81 & 131.6 & 109.09 & 108.41 \\

\([8,\,8,\,8,\,8,\,8,\,8]\)
& 258.09 & 275.84 & 231.34 & 233.92 & 181.14
& 199.82 & 170.79 & 171.15 & 145.68 & 130.85
& 155.91 & 139.5 & 154.91 & 124.78 & 125.51 \\

\([2,\,8,\,4,\,8,\,6,\,5]\)
& 208.94 & 214.15 & 191.64 & 180.52 & 135.35
& 166.58 & 137.52 & 148.78 & 110.97 & 111.87
& 148.74  & 126.34 & 148.34 & 107.64 & 108.82 \\

\([2,\,2,\,2,\,8,\,8,\,8]\)
& 202.39 & 194.64 & 183.65 & 169.09  & 137.2 
& 148.06 & 145.12 & 147.42 & 119.89 & 119.8
& 148.06 & 145.12 & 147.42 & 119.37 & 119.75 \\

\([2,\,8,\,2,\,8,\,2,\,8]\)
& 207.37 & 193.02 & 182.72 & 171.45 & 127.93
& 154.64 & 145.12 & 146.04 & 111.79  & 110.04
& 146.84 & 145.32 & 146.04 & 109.32 & 109.72 \\

\bottomrule
\end{tabular}
\end{table*}

\section{Conclusions}\label{sec:conclusions}
This paper studied joint sensing-data offloading and edge-inference scheduling in multi-UAV networks. A unified communication--computation model was developed by linking the upload completion times of sensing streams to the release times of branch-entry nodes in a multi-branch DNN. Based on this model, an end-to-end latency minimization problem with a synchronization penalty was formulated. To solve this problem, a GA-based joint scheduler, \texttt{GA-Joint}, was proposed to jointly optimize uplink resource allocation, DAG priority, and core mapping. Two lightweight variants, \texttt{GA-DACS} and \texttt{GA-DAG}, were further designed to balance performance and complexity by focusing on communication-side and computation-side optimization, respectively. Simulation results showed that release-time-aware scheduling effectively improves communication--computation overlap and reduces end-to-end latency under diverse system settings. In particular, \texttt{GA-Joint} achieves the best or near-best overall performance, while \texttt{GA-DACS} provides a lower-complexity alternative with near-\texttt{GA-Joint} performance in many cases.

\ifCLASSOPTIONcaptionsoff
  \newpage
\fi

\end{document}